\pdfoutput=1

\documentclass[sigconf,nonacm]{acmart}
\usepackage{siunitx}
\usepackage{amsmath}
\usepackage{graphicx}
\usepackage{dsfont}
\usepackage{booktabs}
\usepackage{array}
\usepackage{booktabs} 
\usepackage{amsfonts}
\usepackage{xspace}
\usepackage{breqn}
\usepackage{hyperref}
\usepackage{caption}
\usepackage{float}
\usepackage{xcolor}
\usepackage{tabularx}
\usepackage{bbm}

\title{A Survey on Large Language Models in Multimodal Recommender Systems} %

\author{Alejo López-Ávila}
\email{vonnegut31416@gmail.com}
\affiliation{
  \institution{}
  \city{}
  \country{}
}

\author{Jinhua Du}
\email{jinhua.du@huawei.com}
\affiliation{
  \institution{Huawei London Research Centre}
  \city{}
  \country{}
}
\newcommand{\GUI}{\ensuremath{\mathbb{G}_{\text{UI}}}~}

\begin{document}

\begin{abstract}
Multimodal recommender systems (MRS) integrate heterogeneous user and item data, such as text, images, and structured information, to enhance recommendation performance. The emergence of large language models (LLMs) introduces new opportunities for MRS by enabling semantic reasoning, in-context learning, and dynamic input handling. Compared to earlier pre-trained language models (PLMs), LLMs offer greater flexibility and generalisation capabilities but also introduce challenges related to scalability and model accessibility. This survey presents a comprehensive review of recent work at the intersection of LLMs and MRS, focusing on prompting strategies, fine-tuning methods, and data adaptation techniques. We propose a novel taxonomy to characterise integration patterns, identify transferable techniques from related recommendation domains, provide an overview of evaluation metrics and datasets, and point to possible future directions. We aim to clarify the emerging role of LLMs in multimodal recommendation and support future research in this rapidly evolving field.
\end{abstract}
\maketitle
\section{Introduction}
\label{sec:introduction}
Multimodal recommender systems aim to integrate diverse user and item information, including text, images, structured data, and user behaviour, into coherent recommendation pipelines. Traditional approaches, such as collaborative filtering and neural encoders, often face persistent challenges like data sparsity, cold-start problems, and modality misalignment. Large language models (LLMs) offer a new paradigm to address these issues through pretrained knowledge, semantic flexibility, and inference-time adaptability.

Unlike earlier pretrained language models (PLMs) such as BERT or RoBERTa, LLMs (e.g., GPT-3, PaLM, LLAMA) provide richer world knowledge and support advanced reasoning and prompt-based conditioning mechanisms. These features enable them to accommodate diverse and evolving recommendation inputs without retraining, which is particularly valuable in cold-start and cross-domain scenarios.

This survey examines how LLMs reshape the design of multimodal recommender systems. We focus on their integration through prompting, training, and data adaptation, as detailed in Section~\ref{sec:llmmethods}. These techniques also interact with fundamental components such as disentanglement, alignment, and fusion, which are central to cross-modal learning in MRS (Section~\ref{sec:mrsmethods}).

We also synthesise recent developments and highlight future research directions (Section~\ref{sec:future}). To support further work, we provide an extensive dataset list (Appendix~\ref{sec:datasets}), a structured classification of evaluation metrics (Appendix~\ref{sec:metrics}), and a comprehensive table of abbreviations (Appendix~\ref{sec:notations}).

\subsection{Strategy}
This survey centres on how LLM-specific capabilities—reasoning, prompting, and modality adaptation—reshape the design of MRS (Section~\ref{sec:taxonomy}). To maintain this focus, we deliberately reduce attention to traditional architectural components, such as modality-specific encoders, which are the main emphasis in prior surveys~\cite{MRS1_survey, MRS2_survey}. Similarly, we do not address the typology of recommender architectures (e.g., GCNs, transformers), as these are comprehensively covered elsewhere~\cite{Gheewala2025, survey_rs_gcns, survey_dl_cf, survey_rs_gnns}.

In contrast, we highlight underexplored modalities such as tabular and numerical data, and examine how they can be integrated into LLMs—a topic often overlooked in previous works. We also include methods from related branches of recommender systems, such as sequential and knowledge-aware recommendation, when their LLM-based techniques are transferable to multimodal settings. This broadens the relevance of our analysis and supports a more comprehensive characterisation of LLM–MRS interactions.

\subsection{Differences among Other Surveys on MRS}
\label{sec:differences}
While previous surveys on MRS have provided extensive reviews, they typically adopt encoder-centric taxonomies focused on architectural components such as modality-specific encoders, fusion mechanisms, or loss functions~\cite{MRS1_survey, MRS2_survey}. In contrast, our work focuses on the transformative impact of LLMs, which extend beyond traditional encoder pipelines.

\paragraph{\textbf{LLMs Beyond Encoders}} Traditional surveys often treat encoders as the central mechanism for modality representation. However, LLM-based models alter this paradigm. LLMs enable flexible input handling through prompt engineering and operate directly over multimodal summaries or structured formats (e.g., JSON, tabular text). These capabilities shift the role of the model from a static encoder–decoder to a dynamic agent capable of contextual reasoning, intent inference, and interaction with external tools or latent spaces.

\paragraph{\textbf{Taxonomy Based on LLM-Specific Functions}} our taxonomy departs from standard architectural categorisations towards LLM-centric integration strategies: prompting (Section~\ref{sec:prompt}), training approaches (Section~\ref{sec:training}), and data type adaptation (Section~\ref{sec:datatypeadaptation}), along with the model’s role in MRS tasks like disentanglement and alignment (Sections~\ref{sec:disentangle} and~\ref{sec:alignment}). These dimensions capture how LLMs enable novel forms of reasoning and cross-modal alignment.

\paragraph{\textbf{Inclusion of Transferable Techniques from Related RS Domains}} Given the recency of LLM adoption in MRS, we provide a broader perspective. We incorporate LLM-based techniques from adjacent RS domains, such as sequential, textual, or knowledge-aware recommendations that introduce transferable mechanisms. This inclusion expands the design space and highlights emerging patterns applicable to multimodal settings.

\subsection{Taxonomy}
\label{sec:taxonomy}
This survey introduces a new taxonomy tailored to the integration of LLMs in multimodal recommender systems (MRS), moving beyond the encoder- or loss-centric classifications adopted in earlier works~\cite{MRS1_survey, MRS2_survey, LLMSRS_survey}. 
The arrival of LLMs introduces a qualitatively different design space centred on reasoning capabilities, prompt-based control, and dynamic adaptation at inference time.

Our taxonomy captures this shift by structuring the reviewed work into three primary categories:
\begin{enumerate}
    \item \textbf{LLM Methods} (\textit{Section~\ref{sec:prompt}–\ref{sec:datatypeadaptation}}): These methods are distinguished by their use of LLM-specific techniques. We subdivide them as follows:
    \begin{itemize}
        \item \textbf{Prompting Techniques (Section~\ref{sec:prompt})}, including hard prompts (e.g., ID-based tokens), soft prompts, hybrid templates, and reasoning prompts.
        \item \textbf{Training Strategies (Section~\ref{sec:training})}, such as fine-tuning (FT), parameter-efficient methods like LoRA and QLoRA, or agents.
        \item \textbf{Data Type Adaptation (Section~\ref{sec:datatypeadaptation})}, covering techniques for adapting non-text modalities—such as image, tabular, or behavioural data—into formats suitable for LLM input via summarisation or structured prompts (e.g., JSON).
    \end{itemize}

    \item \textbf{MRS-Specific Techniques} (\textit{Section~\ref{sec:disentangle}–\ref{sec:fusion}}): These categories reflect long-standing challenges in multimodal recommendation, revisited through the lens of LLMs:
    \begin{itemize}
        \item \textbf{Disentanglement (Section~\ref{sec:disentangle})}, including approaches that separate modality-specific and shared signals via latent factor models, contrastive learning, or variational inference.
        \item \textbf{Alignment (Section~\ref{sec:alignment})}, covering mechanisms that synchronise inputs across modalities or link external knowledge to LLM embeddings.
        \item \textbf{Fusion (Section~\ref{sec:fusion})}, where diverse modalities are combined using early, mid-level, or late fusion strategies, with different fusion techniques.
    \end{itemize}

    \item \textbf{Main Trends and Future Directions} (\textit{Section~\ref{sec:future}}): This final category synthesises emerging patterns in the field, and the evolving role of LLMs as reasoning agents within the recommendation pipeline.
\end{enumerate}

Finally, in Appendix~\ref{sec:metrics}, we provide an extensive and structured overview of both standard and emerging metrics, including NLP-derived metrics such as BLEURT, and LLM-based evaluators, that are increasingly used to assess MRS performance. In Appendix~\ref{sec:datasets}, we present a comprehensive list of multimodal datasets, extending previous surveys with additional domains and modalities.

\subsection{Contributions of the Survey}
\label{sec:contributions}
Surveys on MRS tend to overlook the specific challenges and opportunities introduced by LLMs. For instance, \citet{MRS1_survey} briefly mention Multimodal LLMs only in the context of future directions. Similarly, surveys focusing on PLMs in recommendation systems tend to concentrate on BERT-like architectures, which lack the in-context reasoning and tool-augmented capabilities of more recent LLMs.

This survey fills that gap by offering a comprehensive and LLM-focused analysis of the emerging landscape in multimodal recommendation. We also examine relevant strategies developed in adjacent RS domains that may be transferable to the MRS setting. Our goal is to provide a forward-looking framework that captures the current state and the research frontier. In particular, our contributions are fourfold:
\begin{enumerate}
    \item \textbf{A New Classification Framework for LLMs in MRS:} We propose a novel taxonomy that organises the integration of LLMs in MRS into distinct components—including prompting paradigms, parameter-efficient tuning methods, modality adaptation strategies, and alignment techniques—each treated as a standalone design axis, allowing for a more precise understanding of LLM-specific mechanisms.
    \item \textbf{Cross-Domain Integration and Boundary Extension:} We include relevant methods from neighbouring areas of recommendation research, such as textual or behavioural recommendation, where LLM-based strategies (e.g., summarisation, in-context learning, or reasoning prompts) are present but not yet adapted to multimodal setups, revealing transferable innovations and contextualising LLM–MRS development limitations.
    \item \textbf{Mapping Current Trends and Identifying Gaps:} We offer a synthesis of ongoing developments and underexplored directions across all surveyed works (Section~\ref{sec:future}). This analysis highlights where research is converging and new contributions are still needed.
    \item \textbf{Expanded Resources on Metrics and Datasets:} In contrast to prior surveys, we provide an extended and structured overview of evaluation metrics (Appendix~\ref{sec:metrics}) that reflects both MRS-specific goals (e.g., diversity, novelty, multi-objective trade-offs) and recent LLM-driven practices (e.g., LLM-based evaluation, NLG scores). We also compile a more complete and diverse set of publicly available datasets for multimodal and LLM-based recommendation (Appendix~\ref{sec:datasets}).
\end{enumerate}

\section{LLM methods}
\label{sec:llmmethods}
The appearance of ChatGPT at the end of 2022 marked a turning point in the NLP field. Although large language models (LLMs) such as GPT-3 already existed, the integration of Reinforcement Learning from Human Feedback (RLHF), initially proposed earlier \cite{RLHF}, significantly improved their reasoning and conversational abilities, making them more aligned with human expectations. This leap in capability quickly attracted interest across fields, including recommender systems (RS), where researchers began to explore the potential of LLMs for tasks like search \cite{ischatgptgoodatsearch2023} and fairness-aware recommendation \cite{ischatgptfairforrecommendation2023}.

However, the integration of LLMs into RS workflows also revealed critical limitations. These models require substantial computational resources, and their access is often constrained to black-box input-output APIs. Additionally, the latency introduced by LLM inference can be problematic in recommendation contexts, which typically demand low-latency interactions for real-time user engagement.

A wave of LLM adaptation techniques emerged in 2023 to mitigate these challenges. These include efficient parameter adaptation strategies that avoid full model updates, such as Adapters, LoRA, and QLoRA \cite{Adapters2020, LoRA, QLoRA} and prompting-based methods that steer LLMs through carefully designed inputs. This section categorises these approaches, detailing their use in multimodal recommendation settings and discussing their performance, generality, and computational cost trade-offs.

\subsection{Prompting}
\label{sec:prompt}

In this context, prompting emerges as a central strategy for integrating LLMs in MRS. Prompting enables models to adapt to various tasks without retraining the whole model, and is especially useful in cold-start and multi-task scenarios.

Unlike fine-tuning, prompting does not require updating the internal weights of the LLM, making it a lightweight and interpretable alternative, particularly valuable when working with black-box APIs or restricted compute environments.
Prompting is also cost-effective in real-world scenarios where updating or hosting large-scale LLMs may be infeasible. Techniques such as soft prompting or structured templates enable flexible adaptation across domains without requiring full retraining, thereby reducing infrastructure overhead while maintaining competitive performance.

In MRS, prompting plays various roles: it can inject structured multimodal context into the input stream, act as a mechanism for eliciting preferences or routing tasks, or serve as a reasoning scaffold through decomposition or chaining. This versatility makes prompting an attractive choice across many LLM–MRS applications.

\begin{figure}[h]
    \centering
    \includegraphics[width=0.9\linewidth]{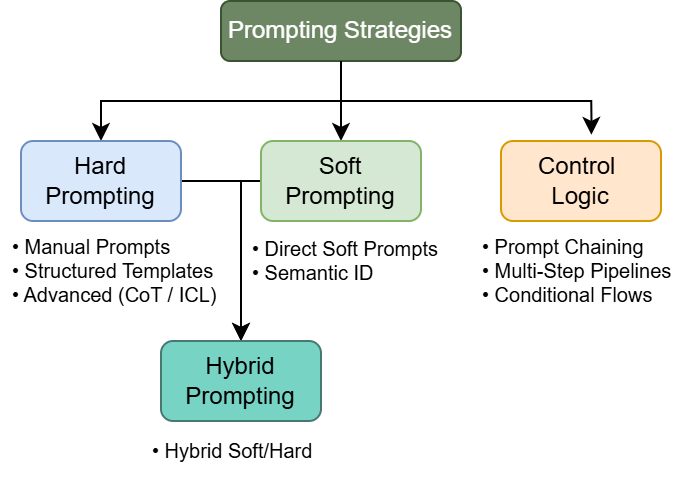}
    \caption{Prompting strategies in LLM-based MRS.}
    \label{fig:prompt_diagram}
\end{figure}

We introduce a classification of prompting methods in \ref{fig:prompt_diagram}.
Agent-based prompting, which extends these ideas to persistent and interactive agents, will be discussed in Section~\ref{sec:agents}.

\subsubsection{Hard prompting}
\label{sec:hard_prompt}
\paragraph{\textbf{Hard prompting}}
Hard prompting (\cite{hardprompt}) is the earliest and most widely adopted prompting paradigm. It involves crafting fixed input templates—manually or programmatically, guiding the LLM's behavior during inference without updating its internal weights. These prompts are human-readable and often designed through domain knowledge, intuition, or task-specific rules. This simplicity allows rapid adaptation across domains, making hard prompting suitable for black-box LLM APIs, resource-constrained environments, or low-data settings.

Despite its practicality, hard prompting is often brittle—small changes in phrasing can significantly affect performance. Moreover, its reliance on human effort limits scalability, especially in complex or multimodal domains. Nevertheless, a variety of techniques have emerged under this paradigm, including manual prompt engineering \cite{0-shotRec2023, IntellectSeeker, LLAMA_E, LLLM_Rec, RLMRec2024}, rule-based structured templates \cite{CTRL2023, LLM4KGC2023, PIE, LLM_Rec, Mender, HKFR}, and reasoning-oriented formats like in-context learning and chain-of-thought prompting \cite{ViML, GIRL, Multi_LLM_GNSS}. In this section, we classify these approaches by design strategy and analyse their roles in LLM-based multimodal recommender systems.

\textbf{Manual Prompt Engineering:}
This approach relies on manually crafted prompt patterns based on intuition, domain expertise, or trial-and-error, typically without any learned structure. These prompts are human-designed and task-specific, requiring no updates to model parameters. One of the earliest and most representative works is \citet{0-shotRec2023}, which introduces a zero-shot recommendation method using hard prompts phrased as movie lists (e.g., “Movies like Matrix, Inception, <m>”) and scores candidates based on LLM likelihoods.

Manual prompts can also serve as a base for more complex designs: \citet{IntellectSeeker} uses prompt-based academic concept substitution, while \citet{LGIR} guides resume completion through user self-descriptions or interaction data. In more technical domains, \citet{Multi_LLM_GNSS} apply handcrafted prompts to interpret GNSS snapshots and associated metadata. Some works, like \citet{LLAMA_E} and \citet{LLLM_Rec}, begin with manually designed prompts but extend them with fine-tuning strategies such as instruction tuning or LoRA—these will be revisited in Section~\ref{sec:fine_tuning_adapters}.

These examples illustrate the flexibility of manual prompting across domains, though their performance is sensitive to phrasing and requires considerable design effort.

\textbf{Structured template prompts:} They are built using systematic, rule-based formats that organise information into consistent and interpretable patterns. This enables automation, reduces variability, and allows easy adaptation to different domains. Such templates often separate user and item fields using special punctuation or semantic cues and are particularly suited to converting tabular, metadata, or sequential data into natural language inputs.

Early examples focus on transforming structured input into textual descriptions: CTRL \citet{CTRL2023} use handcrafted templates to encode user profiles and item features (e.g., “This is a user, gender is female, age is 18, who has recently watched Titanic|Avatar. This is a movie, title is The Terminator,...”) to train LLMs on user-item interactions. 
LLM4KGC \citet{LLM4KGC2023} refine template-based prompting with few-shot ICL examples, emphasising clarity, relevance, and consistency in format.
Similarly, PIE \citet{PIE} explores structured zero-shot prompts with explicit task descriptions and format constraints for product information extraction, 
demonstrating gains when few-shot ICL is added for ambiguous cases.

Other methods increase complexity or modularity in prompting. For example, \citet{KAR} introduce \emph{factorisation prompting}, which decomposes recommendation reasoning into smaller subtasks—each encoded via a separate sub-prompt. This modular approach improves interpretability and enables better task generalisation.

A third group of works focuses on deriving prompts directly from user behaviours or historical data. \citet{LLM_Rec} utilizes summarisation, paraphrasing, recommendation-driven, and engagement-guided prompts to enhance item description generation; \citet{HKFR} encodes heterogeneous, multi-type user behaviour logs into unified text templates; and \citet{Mender} designs per-timestep prompts based on sequential interaction histories and reviews.

Beyond individual prompting strategies, some works analyze the broader implications of prompt design choices. \citet{bias} conducts a comparative evaluation of different hard prompting strategies, such as accuracy-focused prompts, diversification prompts, and reasoning-based formats like CoT and ICL, in the context of provider fairness and diversity. 

\textbf{Reasoning-Based Hard Prompts:} This category comprises prompts designed to support multi-step reasoning or inference, extending beyond surface-level formatting. These prompts often utilise CoT logic or ICL and can help LLMs perform more structured analysis, even without any parameter updates.

ViML \cite{ViML} uses BLOOM-176B \cite{bloom} to generate natural language music descriptions from tags and sparse human annotations. Among their prompting techniques are an ICL-based method that combines manually crafted prompts with examples, and Categorical Data Integration with templates and rephrasing, which enables a richer understanding of music.
GIRL \cite{GIRL} reformulates job recommendation as a text generation task, using structured prompts that concatenate candidate and job descriptions data. Similar in spirit to CoT, leverages prompt-based task reformulation and textual augmentation, followed by supervised fine-tuning and reward-based alignment.
Multi-LLM-GNSS \cite{Multi_LLM_GNSS} also employs a reasoning-based strategy, combining handwritten task prompts with retrieval-augmented ICL to guide the VLM LLaVA \cite{llava} in the multimodal interpretation of GNSS snapshots and textual. Notably, this work spans both manual prompting and reasoning, without any parameter tuning in complex scenarios.

These methods demonstrate how reasoning-oriented prompts can improve the LLM’s ability to navigate more cognitively demanding recommendation tasks.

\subsubsection{Soft prompting}
\label{sec:soft_prompt}
\paragraph{\textbf{Soft prompting}}
Soft prompting is a parameter-efficient adaptation technique for PLMs and LLMs, where trainable continuous embeddings (soft tokens) are prepended to the input, replacing or complementing textual prompts. Unlike hard prompting, soft prompts are not human-readable and require access to the model’s embedding space and training loop, making them less interpretable but more adaptable to domain-specific tasks. These approaches are beneficial in scenarios where full fine-tuning is costly, infeasible, or undesirable due to constraints on compute or access to model weights.

This strategy enables task-specific adaptation while keeping the base LLM frozen, making it particularly valuable for integrating structured or multimodal data. Soft prompting typically requires some supervision, with embeddings optimised over example inputs. We distinguish two key subtypes relevant for MRS: direct soft prompting and semantic ID prompting.

\textbf{Direct soft prompting} injects learnable continuous embeddings (soft tokens) into the input space of a frozen LLM, without requiring explicit text or structured prompt design. Techniques such as Prefix-Tuning \cite{PrefixTuning}, Prompt Tuning \cite{PromptTuning}, and P-Tuning \cite{PTuning, PTuningv1, PTuningv2} fall into this category. These approaches can be tailored per input type or modality.

An example of this approach in MRS is found in TMF \cite{TripletFusion2024}, which combines instruction-tuned textual input with soft prompts. Specifically, it extracts a vector representation of user behaviour from a \GUI interaction graph, projects it via a feedforward network, and appends it as a soft embedding after the user's textual description. Similarly, item representations from \GUI are fused with image and text features through encoders, with the result appended as another soft prompt following the item's textual description. PromptMM \cite{promptmm} introduces soft prompt-tuning in the teacher model to bridge the semantic gap between multi-modal inputs and collaborative signals, enabling task-adaptive guidance for knowledge distillation.

A second subtype, particularly relevant in recommendation settings, focuses on representing user and item IDs as learnable soft prompts.

\textbf{Semantic ID prompting} replaces user/item IDs with learned continuous embeddings that act as soft prompts, bypassing the need for tokenisation or natural language formatting. This approach is particularly relevant in recommender systems where entities are often represented as unique identifiers.

Several works adopt this strategy. For example, VIP5 \cite{VIP5} incorporates semantic embeddings for user and item IDs alongside multimodal features such as CLIP-derived image tokens. PEPLER \cite{PersonalizedPromptLearning} proposes soft prompt injection for user and item ID vectors directly into a frozen LLM, thereby avoiding the need to map these IDs to natural language. Related works \cite{PAD2024, TTDS2024} adopt similar strategies, often employing Sequential Tuning where the LLM is first frozen while the prompts are tuned, and then both are updated jointly.

These methods demonstrate how soft prompting can flexibly encode structured or user-specific context, providing a lightweight and modular approach to inject structured or user-specific context (see also \cite{IDRec_vs_MoRec} for a comparison between ID-based and modality-based item representations, showing that modality encoders can match ID embedding performance even without fine-tuning).

\subsubsection{{Hybrid Techniques}}
\label{sec:hybrid_prompt}
\textbf{Hybrid prompting} combines fixed hard prompts with learnable soft components, balancing interpretability and adaptability. These techniques are particularly effective when integrating structured or multimodal data, enabling task-specific customisation without sacrificing generalisation. Hybrid prompting is especially valuable in cold-start and personalisation scenarios, where both fixed patterns and learnable user/item-specific signals are required.

Several works explicitly combine hard and soft prompting to balance fixed logic with learnable personalisation. For example, TMF \cite{TripletFusion2024} integrates Instruction Tuning (IT) with soft prompts, enabling the combination of structured text with learned embeddings. Similarly, PEPLER \cite{PersonalizedPromptLearning} introduces a hybrid framework that fuses manually constructed prompts with soft prompt learning tailored for integrating user and item IDs into pre-trained LLMs. For the hard component, it proposes PEPLER-D, which maps user and item IDs to domain-specific features (e.g., movie titles or attributes), aligning them with the semantic space of models like BERT \cite{BERT} or GPT-$2$ \cite{GPT2}. The resulting fixed-size prompts serve as interpretable explanations without modifying the base model.

Another hybrid approach is EUIRec (DUIP) \cite{EUIRec}, which dynamically generates soft prompts from an LSTM’s hidden state to capture current user intent, while using static hard prompts to encode historical preferences. These components are fused into a single input and passed to the LLM for next-item prediction, effectively bridging short-term and long-term behaviour representations.

Other approaches build hybrid prompts from multimodal features and lightweight modules such as adapters. Although not always labelled explicitly, these methods combine fixed structure with learnable embeddings, making them hybrid by design. PROMO \cite{PROMO} introduces a soft prompting framework tailored for cold-start item recommendation. Rather than relying on full-model fine-tuning, it encodes pinnacle feedback—high-value positive user interactions into learnable prompts through a personalised prompt network. These prompts are optimised to mitigate bias toward warm-start items and improve personalisation with minimal prompt parameter updates. 

VIP5 \cite{VIP5} constructs multimodal personalised prompts that include textual fields, semantic user and item IDs, and visual tokens derived from CLIP image features. A projection network converts the visual input into token-level embeddings inserted into the LLM's input stream. The backbone MLLM remains frozen, while adapters are trained to enable flexible integration of these modalities.
UniMP \cite{UniMP} interleaves text, structured metadata, and visual embeddings from items into a flattened input sequence, forming a hybrid prompt for a generative LLM. Visual inputs are introduced through the placeholder [IMG] from a learned cross-attention fusion module rather than tokenisation, creating a soft representation conditioned on the image content and aligned with other item attributes.

While not directly prompting an LLM, some hybrid-style systems use learned prompts for selecting specialised agents. AgentRec (\cite{AgentRec}) employs standardised hard prompts (via rephrase-and-respond templates) to convert user inputs into sentence embeddings, used for nearest-neighbour retrieval over a pre-encoded corpus of agent prompts. These hard prompts serve as inputs to a Sentence-BERT encoder fine-tuned to cluster task types by agent specialisation. The retrieved agent is then activated to handle the downstream task.

These hybrid prompting strategies demonstrate how LLM-based MRS systems can flexibly blend static templates with dynamic embeddings, enabling generalisation, personalisation, and scalable multimodal reasoning.

\subsubsection{Control Logic}
\label{sec:control_logic}
\paragraph{\textbf{Control logic prompting}} refers to the use of structured, programmatic strategies to coordinate multiple LLM calls, often arranged in multi-step, branching, or conditional flows. Unlike single-shot prompting, control logic enables the LLM to reason through complex tasks by decomposing them into sequential subtasks or navigating decision graphs. This approach is especially effective in settings that demand interpretability, modular reasoning, or adaptive execution, such as recommendation workflows involving feature selection, structured data extraction, or diagnostic inference. In multimodal recommender systems (MRS), control logic supports the orchestration of LLMs across heterogeneous inputs and operations, offering a flexible means to embed domain knowledge or logic-based control while maintaining the expressiveness of language models.

Several works employ control logic through \textbf{multi-stage prompting} pipelines, where the task is broken down into discrete reasoning stages—such as extraction, ranking, validation, or synthesis—executed sequentially via LLM interactions. This decomposition enables modular, interpretable workflows and progressive refinement of outputs.
AltFS \cite{AltFS} applies such a pipeline to the feature selection task in recommender systems. It combines LLM-based ranking (via semantic and contextual relevance) with a bridge network that refines the selected features during model training. The system iteratively prompts the LLM to evaluate candidate features, yielding a lightweight and generalisable selection mechanism that incorporates both world knowledge and task-specific context.

GraphJudger (\cite{GraphJudger}), not an RS approach but a KG approach that can be applied to our setting (Section \ref{sec:kgconversion}), uses prompt-driven control logic to coordinate multiple LLM calls across a multi-phase KG construction pipeline. In phase one, an LLM iteratively drafts and revises a knowledge graph from unstructured input using guided prompts. At phase two, a fine-tuned LLaMA model performs triple-level fact-checking, treated as a sequence-to-sequence task, and is prompted to refine the KG.

Other methods use \textbf{tree- or graph-based control flows}, where LLMs operate within structured decision graphs, such as medical diagnosis trees or knowledge graph refinement paths. These approaches support hierarchical decision-making by prompting over intermediate nodes and updating outputs based on prior steps. IndustryScopeGPT~\cite{IndustryScopeGPT2024} integrates a pre-built multimodal knowledge graph (KG) and applies Monte Carlo Tree Search (MCTS) to structure LLM interactions. The system formulates decision-making as a tree search, where each node represents a prompting step over structured data. Guided by an enhanced UCT algorithm, the method iteratively selects actions, expands the tree, and back-propagates feedback from LLM assessments to refine outputs, ultimately selecting the most optimal trajectory.

HiRMed \cite{HiRMed} introduces a tree-structured medical recommendation framework, where each node contains a specialised RAG-based reasoning module that performs localised analysis. The system adapts its reasoning dynamically according to medical urgency and diagnostic uncertainty, simulating step-wise expert decision-making. NegGen \cite{NegGen} applies a multi-phase prompting pipeline to generate contrastive negative samples for multimodal recommendation with an MLLM. Its sequential logic includes image-to-text generation, attribute masking, and guided text completion, to synthesise hard negatives. This logic ensures the negatives are semantically plausible yet distinct.

While some of these approaches resemble early forms of agent interaction, they do not maintain a persistent state or autonomy. We defer detailed discussion of agent-based prompting to Section~\ref{sec:agents}.
\subsection{Training}
\label{sec:training}
Training or adapting LLMs for recommender systems presents challenges due to their scale and the limited availability of domain-specific multimodal pretraining. Most works avoid full pretraining from scratch, instead opting for parameter-efficient fine-tuning, black-box prompting, or leveraging pretrained multimodal LLMs (MLLMs). The only LLM pretrained from scratch specifically for recommendation is P5~\cite{P52023}, which extends the T5~\cite{T5} framework by converting multiple tasks, such as rating, review generation, sequential recommendation, and zero-shot predictions, into unified text-to-text formats. However, although P5 incorporates scores, IDs, and text descriptions, it is not a multimodal model. 

\begin{figure*}[t]
    \centering
    \includegraphics[width=0.85\textwidth]{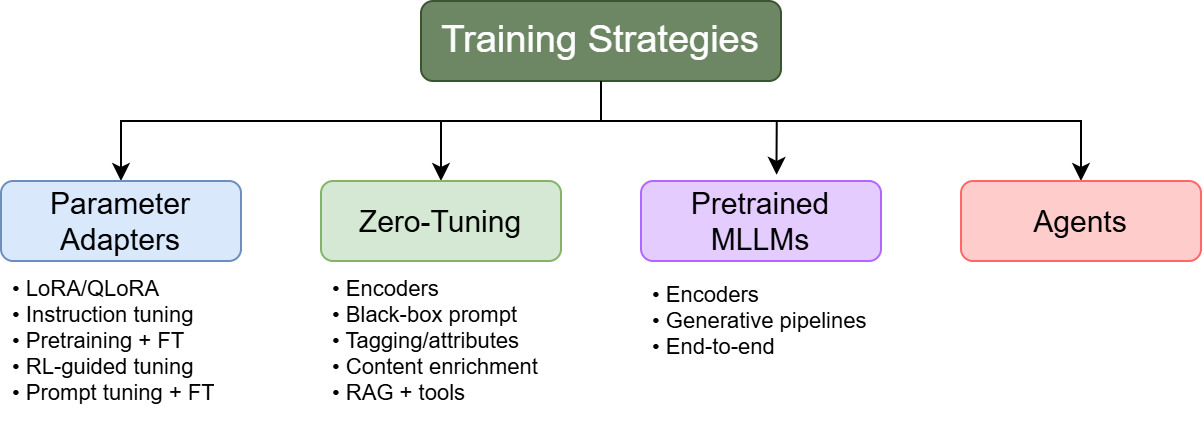}
    \caption{Training strategies for adapting LLMs in Multimodal Recommender Systems.}
    \label{fig:training_diagram}
\end{figure*}

In this section, we review how LLMs are adapted or used for downstream recommendation tasks across four main categories: (i) parameter adaptation, (ii) zero-tuning usage, (iii) pretrained MLLMs, and (iv) agent-based systems (see \ref{fig:training_diagram}).

\subsubsection{Parameter Adaptation}
\label{sec:fine_tuning_adapters}
Parameter adaptation refers to strategies that modify a subset of the LLM's parameters or an external complementary network to specialise it for downstream recommendation tasks. This includes full fine-tuning, lightweight adapter methods such as LoRA~\cite{LoRA} and QLoRA~\cite{QLoRA}, and more complex pipelines that combine pretraining and fine-tuning stages. These approaches offer high flexibility and allow model behaviour to be shaped explicitly for task needs, but they require access to the model weights and computational resources. Within this category, we distinguish five main groups: adapter-based and lightweight tuning, instruction or task-specific fine-tuning, pretraining–fine-tuning pipelines, reinforcement learning–augmented tuning, and prompt-enhanced tuning (combining structured or soft prompts with fine-tuning).

\paragraph{\textbf{Adapter-based and Lightweight Fine-tuning}}
A popular direction involves lightweight adaptation of LLMs or modality encoders through \textbf{LoRA or adapters}. LLAMA-E~\cite{LLAMA_E}, LGIR~\cite{LGIR}, and LLMSeqPrompt from LLLM-Rec~\cite{LLLM_Rec} apply LoRA for domain-specific or sequential recommendation tasks, while Mender~\cite{Mender} uses LoRA for fine-tuning multimodal LLMs. iLoRA~\cite{iLoRA} extends this approach with a gating mechanism that dynamically selects expert participation during adaptation.

\paragraph{\textbf{Instruction and Task-specific Fine-tuning}}
Other works perform task-specific fine-tuning, often guided by instruction data or downstream task structure. Multi-LLM-GNSS~\cite{Multi_LLM_GNSS} fine-tunes an MLLM with multi-turn instruction data to support visual-logical reasoning in GNSS interference monitoring. GIRL~\cite{GIRL} applies Supervised Fine-Tuning (SFT) and further refines the model through Reinforcement Learning from Reflective Feedback (RLRF). Other works include RecInDial~\cite{RecInDial}, which introduces KG information through a GCN (see Section~\ref{sec:kgconversion}), and ICPC~\cite{ICPC}, which adapts LLMs to generate names for clustered user journeys. These models tailor LLM behaviour using structured downstream signals or annotations, and frequently incorporate auxiliary components for task grounding.

\paragraph{\textbf{Pretraining–Fine-tuning Pipelines}}
Several multimodal approaches adopt a pretraining–fine-tuning pipeline to improve cross-modal generalisation.
MISSRec~\cite{MISSRec} uses modality-specific adapters and fine-tunes only adapters and item embeddings for efficient adaptation. PMMRec~\cite{PMMRec} combines multimodal contrastive, denoising, and alignment pretraining with full model fine-tuning in downstream domains.

\paragraph{\textbf{RL-augmented Tuning}}
Some models integrate LLMs into reinforcement learning pipelines, fine-tuning them as part of the policy learning stage. 
iALP~\cite{iALP} uses an LLM to generate user preferences via structured prompts, which are then used to pre-train a reinforcement learning policy offline using an actor-critic architecture. The model undergoes a second stage of training where it is either directly fine-tuned with online feedback (A-iALP\textsubscript{ft}) or adapted through a hybrid policy mechanism that gradually replaces the frozen policy (A-iALP\textsubscript{ap}). The LLM is fine-tuned as part of the reward and action modelling during offline training, playing an active role in shaping the initial policy behaviour. A different use of RL appears in GIRL~\cite{GIRL}, which refines a job description generator through PPO-based reinforcement learning aligned with recruiter feedback.

\paragraph{\textbf{Prompt-enhanced Tuning}} 
Rather than altering model weights, prompt-based adaptation modifies the input interface to elicit desired behaviour from frozen LLMs. In some cases, these prompts are tuned or extended with lightweight fine-tuning. PEPLER~\cite{PersonalizedPromptLearning} introduces a sequential tuning scheme where soft prompts encoding user and item semantics are first trained with the LLM frozen, followed by full fine-tuning for personalisation. TMF~\cite{TripletFusion2024} adapts soft prompts using LoRA to fuse multiple modality vectors into coherent input representations for recommendation. Other works, such as HLLM~\cite{HLLM2024} and HKFR~\cite{HKFR}, also combine hard or structured prompts with LoRA-based tuning to incorporate user preferences or heterogeneous behaviour. These works highlight the flexibility of prompt tuning techniques and their role in low-cost model adaptation, particularly when full retraining is impractical.

\subsubsection{Zero-Tuning Usage}
\label{sec:blackbox}
When full model adaptation is infeasible due to computational or access limitations, many works adopt zero-tuning strategies—using LLMs or modality encoders in a frozen state. This includes extracting fixed representations via encoder models or interacting with black-box LLMs via prompting to produce user profiles, content summaries, or other structured outputs. These approaches are efficient and easy to deploy, but typically lack strong task-specific customisation. We group them into five usage patterns: encoder-based representations, black-box prompting for profile generation, tagging or attribute extraction, content enrichment, and retrieval-augmented generation (RAG)-based or tool-integrated pipelines.

\paragraph{\textbf{Encoder-based Representations}} 
A common strategy when fine-tuning is impractical is to use frozen LLMs or other pretrained encoders to extract fixed vector representations. UniMP~\cite{UniMP} combines frozen modality-specific encoders (e.g., BERT, ViT) with lightweight fusion layers trained across multiple tasks, illustrating how frozen models can support cross-modal personalisation. PAD~\cite{PAD2024} applies LLM2Vect~\cite{llm2vec} to convert decoder-based LLMs into encoders for alignment tasks, a useful adaptation for downstream RS modules. Other works, such as~\cite{IDRec_vs_MoRec}, show that frozen modality encoders (MoRec) can match traditional ID-based methods without any tuning, highlighting the practicality of encoder-based zero-tuning.

\paragraph{\textbf{Black-box Prompting for Profile Generation}} 
Another common zero-tuning setup involves generating intermediate user or item profiles via LLM prompting, treating the model as a black box. KAR~\cite{KAR} generates textual knowledge about users and items using structured factorisation prompts, later processed by a hybrid expert system for recommendation. In EUIRec~\cite{EUIRec}, the LLM output is fused with an LSTM that models user sequences, blending black-box generation with classic architectures. Other works, such as HLLM~\cite{HLLM2024}, use MLLMs to generate updated preferences per session step, feeding these into downstream classifiers.

\paragraph{\textbf{Black-box Tagging/Attribute Extraction}} 
Several approaches rely on LLMs to extract tags or item attributes without any fine-tuning, using zero-shot or few-shot prompting. TagGPT~\cite{TagGPT} exemplifies this direction by integrating OCR and ASR pipelines with black-box LLMs (e.g., GPT, PaLM) for tag generation from multimodal input. PIE~\cite{PIE} applies ChatGPT in zero-shot and ICL setups to extract attribute-value pairs for product enrichment. AltFS~\cite{AltFS} uses prompting to guide iterative feature ranking, with its outputs refined via a bridge network.

\paragraph{\textbf{Black-box Content Enrichment for RS Input}} 
Other methods employ black-box LLMs to enrich or generate descriptive item or node content, improving downstream recommendation models. LLM-Rec~\cite{LLM_Rec} augments item text descriptions using LLMs like GPT-3.5, feeding the enriched inputs into a general RS model. TANS~\cite{TANS} applies prompting on GNN-derived summaries to produce interpretable graph node descriptions, useful for text-free or sparse-text graphs. Other examples include LLM4KGC~\cite{LLM4KGC2023}, which applies few-shot prompting to generate product relation labels in e-commerce KGs.

\paragraph{\textbf{Black-box RAG and External Tool Use}} 
A final group of approaches use black-box LLMs within retrieval-augmented pipelines, often integrating embeddings or graph-based tools externally. K-RagRec~\cite{KRagRec} combines PLM and GNN-based semantic indexing with adaptive KG subgraph retrieval, using LLMs to interpret the retrieved knowledge. HiRMed~\cite{HiRMed} exemplifies an agent-like medical RAG system combining OpenAI-based retrieval, GPT-style reasoning, and a fine-tuned LLaMA module for medical test recommendation. Two approaches in LLLM-Rec \cite{LLLM_Rec} apply frozen OpenAI embeddings for item similarity scoring or BERT4Rec embeddings initialisation.

\subsubsection{Pretrained MLLMs}
MLLMs are increasingly used to process textual, visual, and structured data jointly. Typically used in a frozen or lightly fine-tuned state, they offer strong zero- and few-shot generalisation, making them attractive for multimodal recommendation settings. However, their inference latency can be a bottleneck. In this section, we group MLLM-based approaches into three categories: (i) multimodal encoding for representation learning, (ii) generative pipelines using structured prompts, and (iii) end-to-end architectures embedding MLLMs as central modules.

\paragraph{\textbf{Multimodal Encoding via Pretrained MLLMs}}  
Several works use frozen or lightly tuned MLLMs as flexible encoders to integrate multimodal content. VIP5~\cite{VIP5} injects CLIP-derived visual features into a frozen MLLM using adapters, enabling token-level fusions. UniMP~\cite{UniMP} encodes text, metadata, and image features via frozen modality-specific encoders, then fuses them into a prompt-like input where image embeddings are projected through learned fusion layers before being passed to a generative language model. Other works, such as HLLM~\cite{HLLM2024}, TMF~\cite{TripletFusion2024}, and Mender~\cite{Mender}, apply frozen or lightly tuned MLLMs to fuse multimodal signals via summarisation, soft prompt fusion, or semantic ID generation.

\paragraph{\textbf{MLLM-enabled Generative Pipelines}}  
Some methods leverage MLLMs in structured prompting pipelines for data generation. NegGen~\cite{NegGen} constructs a multi-step generation framework that synthesises contrastive negative image-text examples to enrich training data. SampleLLM~\cite{SampleLLM} uses structured prompts and ICL to synthesise tabular and text data.

\paragraph{\textbf{End-to-End Architectures Leveraging Pretrained MLLMs}}  
A few systems embed MLLMs directly in end-to-end architectures to model user-item dynamics. MM-Rec~\cite{MM_Rec} integrates a pretrained ViLBERT module to jointly encode news titles and image ROIs through hard prompts, combining outputs with co-attention and attention layers for user-news matching. Molar~\cite{Molar} decouples item and user modelling. It uses a fine-tuned MLLM for multimodal item modelling and augments it with a dedicated module to track user behaviour over sequences.

\subsubsection{Agents}
\label{sec:agents}
Agent-based approaches represent an emerging trend where LLMs are instantiated as autonomous agents capable of reasoning, simulating users, or interacting with environments. These systems typically involve planning, tool use, and reflection, and in some cases, adopt multi-agent coordination for more complex tasks. Agents are usually implemented using hard prompts or few-shot examples and may operate iteratively or through orchestration with other models. While still developing, agent-based models show promise for improving personalisation, interpretability, and controllability in recommendations.

Agents reduce work to create the framework and systematic form of formulating hard prompts, like in IndustryScopeGPT (\cite{IndustryScopeGPT2024}) where they concentrate on creating a four-step MCTS using hard-prompt techniques. RecSysLLMsP (\cite{RecSysLLMsP}) uses structured hard prompts to instantiate LLM-based agents (AgentPrompts) with diverse personality traits and evolving preferences. These agents simulate user behavior in different recommendation scenarios (Plurality, Balanced, Similarity) within a social network. LLMs remain frozen and act as decision-making agents to study engagement and polarisation effects.

Agents can be used to adapt different data types, as in SampleLLM (\cite{SampleLLM}), which combines structured CoT prompting with exemplar-based in-context learning to guide LLMs in tabular data synthesis. The instruction prompts are selected from a set of manually designed candidates and iteratively refined via CoT reasoning using the LLM itself. This multi-stage process is coupled with a downstream feature attribution-based control step to reduce distribution bias. Recent works like AgentRec \cite{AgentRec} show that recommendations can now target LLM agents themselves, selecting the best agent for a given prompt. This reflects a shift toward modular and orchestrated LLM systems, where agent selection is key.

\subsection{Data type adaptation}
\label{sec:datatypeadaptation}
\begin{figure*}[t]
    \centering
    \includegraphics[width=0.92\textwidth]{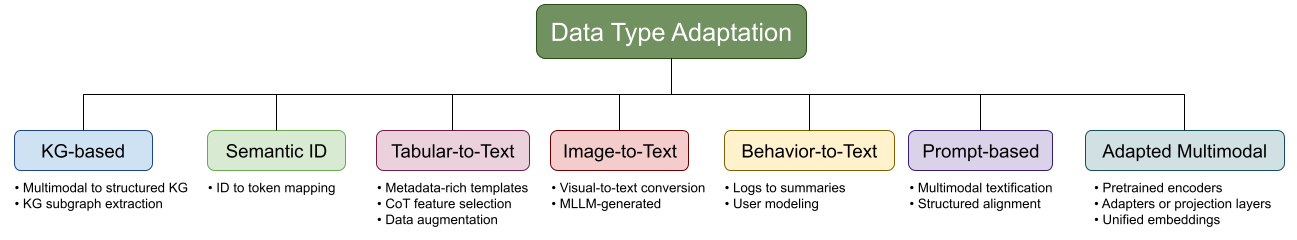}
    \caption{Classification of modality adaptation strategies for integrating structured and multimodal inputs into LLMs.}
    \label{fig:adaptation_diagram}
\end{figure*}

To integrate diverse recommendation signals into LLMs, existing approaches adapt non-textual modalities into LLM-compatible text-like representations. These include structured conversions (e.g., graphs, tabular data), summarisation techniques (e.g., for images or behaviours), and direct multimodal prompting. Adaptation enables LLMs to reason over heterogeneous inputs without requiring architectural changes, and supports both zero-shot prompting and fine-tuning strategies. However, such conversions may introduce alignment errors or information loss, especially when compressing rich modalities into text. Costs vary: KG construction and adapter-based fusion often demand high engineering effort, while prompt-based strategies offer lighter alternatives. We group current methods into eight categories (\ref{fig:adaptation_diagram}): (1) KG-based Conversion, (2) Semantic ID Conversion, (3) Tabular-to-Text Conversion, (4) Image Summarisation for LLM Input, (5) Behavior-to-Text Summarisation, (6) Prompt-based Fusion, (7) Adapted Multimodal Fusion, and will do a final mention for Code-like Structural Conversion.

\subsubsection{KG-based Conversion}
\label{sec:kgconversion}
\paragraph{\textbf{KG conversion}}
Knowledge Graphs (KGs) offer a structured way to unify heterogeneous data, text, images, tabular, or spatial—making them a powerful adaptation layer between multimodal inputs and LLMs. This strategy is particularly valuable in MRS, where rich semantic relations can be leveraged for reasoning, personalisation, or planning. Key benefits include hallucination mitigation and enhanced interpretability, though challenges remain in high-quality KG construction and domain adaptation.

Among the most comprehensive examples, IndustryScopeGPT \cite{IndustryScopeGPT2024} constructs a multimodal KG from geospatial, economic, and textual data, and integrates it with LLMs via recursive Cypher-based queries and in-context learning, outperforming simpler prompt-based methods. RecInDial~\cite{RecInDial} links user dialogue to a KG (DBpedia) for conversational recommendation, encoding subgraphs through relational GCNs and biasing generation via a pointer network.

K-RagRec~\cite{KRagRec} converts unstructured medical data into structured KG representations by extracting entity–relation triples and relevant subgraphs. It combines PLMs for semantic triple extraction with GNNs for encoding structural dependencies, forming a knowledge-enhanced input representation that supports LLM-based recommendation tasks. TANS~\cite{TANS}, while targeting graph classification, offers a graph-to-text conversion strategy using topology-aware prompts, which may inform KG summarisation in RS settings.

Other methods related but not into RS are ADKGD~\cite{ADKGD} (KG from multimodal anomaly logs), and GraphJudger~\cite{GraphJudger} (LLM-assisted KG extraction and refinement), offer a multi-stage strategy for KG construction that could inform robust pipelines in MRS. Or RS papers that use KG to provide structure (\cite{CoWPiRec}).

\subsubsection{Semantic ID Conversion}
\label{sec:semantic_ID_conversion}
\paragraph{\textbf{Semantic ID Conversion}} Maps user/item IDs into semantic token spaces via natural language prompts or learned embeddings. Enhances generalisation but requires careful alignment with collaborative signals.

PEPLER~\cite{PersonalizedPromptLearning} converts user and item IDs into either natural language tokens (discrete prompts) or learned embeddings (continuous prompts), enabling integration into LLMs for explanation generation. Discrete variants use feature-based textual descriptions, while continuous ones are optimised via multi-task regularisation and sequential tuning. 
TTDS~\cite{TTDS2024} introduces vector-quantised semantic indexing for users and items, extending the LLM vocabulary with hierarchical token sets. It jointly learns semantic tokenisation and recommendation through a dual-modality variational autoencoder and fine-tuned prompt-based generation.

These methods provide a flexible interface for integrating ID-based signals into LLMs, laying the groundwork for multimodal extensions where IDs are fused with visual or textual inputs. They also offer promising paths for cold-start handling and fine-grained personalisation in MRS.

\subsubsection{Tabular-to-Text Conversion}
\label{sec:tabular_to_text}
\paragraph{\textbf{Tabular-to-Text Conversion}} Converts structured tabular data into natural language prompts using templates, ICL, or feature-based strategies. These methods enable semantic alignment and lightweight integration with LLMs.

CTRL~\cite{CTRL2023} uses handcrafted templates to convert tabular user and item features into textual prompts aligned with PLM input formats. This conversion supports lightweight models and enhances alignment between semantic and collaborative signals via cross-modal contrastive learning. AltFS~\cite{AltFS} extends this idea by generating metadata-rich prompts for feature selection. It embeds categorical attributes into descriptive templates and leverages ICL for ranking, integrating semantic knowledge with downstream training through a bridge network. SampleLLM~\cite{SampleLLM} introduces a two-stage framework for synthesising tabular recommendation data. It combines Chain-of-Thought prompting with clustering-based example selection and aligns synthetic samples via feature attribution-based importance sampling. PTab~\cite{PTab2022} textualises tabular data through a modality transformation step, enabling PLMs to model structured records using standard fine-tuning.

These approaches demonstrate how templated and reasoning-aware conversions allow LLMs to process structured tabular inputs, with applications in feature selection, data augmentation, and cold-start scenarios.

\subsubsection{Image Summarisation for LLM Input}
\label{sec:image_summarisation}
\paragraph{\textbf{Image Summarisation for LLM Input}} To align image data with text-only LLMs, these approaches convert visual inputs into textual summaries or descriptions using MLLMs or vision-language models. This enables compatibility with generative models without architectural changes. However, this conversion may omit fine-grained visual signals, and its effectiveness depends on the summarisation model's quality.

HLLM~\cite{HLLM2024} addresses the challenge of sequential multimodal recommendation by summarising image sequences into textual descriptions using prompt-based generation, which are then combined with textual item features for user preference modelling. The approach improves scalability by avoiding direct image sequence input into MLLMs and supports fine-tuned prediction via supervised learning. Molar~\cite{Molar} adopts a more integrated strategy, converting item-level multimodal features—including images, structured attributes, and text—into unified embeddings using prompt-based alignment. Structured attributes are textualised, and contrastive loss with temporal modelling enhances alignment across modalities.

These works illustrate how summarisation enables LLM compatibility with image content, laying the groundwork for multimodal fusion via textual representations. The use of MLLMs, their cost and infrequency can be seen as disadvantages of these methods.

\subsubsection{Behavior-to-Text Summarisation}
\label{sec:behaviour_adaptation}
\paragraph{\textbf{Behaviour-to-Text Summarisation}}: These methods transform structured user interaction logs, such as clicks, views, or sequences, into natural language summaries. This enables downstream reasoning, personalisation, and preference modelling within LLMs. While effectively aligning user history with generative models, these methods often rely on handcrafted templates or summarisation quality.

ICPC~\cite{ICPC} clusters semantically related item sequences into "interest journeys" using salient term embeddings, then describes these clusters with natural language summaries generated by LLMs (e.g., PaLM, LaMDA). Prompt tuning and fine-tuning improve fluency and coherence, enabling the model to interpret long behavioural histories. HKFR~\cite{HKFR} encodes heterogeneous behavioural logs (clicks, orders, context) into templated natural language, which is fused into user summaries via ChatGPT. These summaries serve as training data for instruction tuning, allowing the LLM to better capture intent and personalisation cues.

Together, these works show that converting behavioural sequences into textual summaries improves LLM interpretability and user modelling, offering a flexible bridge between structured interaction data and generative interfaces.

\subsubsection{Prompt-based Fusion}
\label{sec:prompt_adaptation}
\paragraph{\textbf{Prompt-based Fusion}}: These methods align heterogeneous inputs via prompting—using ICL, CoT, or templates—to convert multimodal signals into LLM-compatible text. They support zero/few-shot generalisation and avoid model retraining, but can be brittle and template-sensitive.

P5~\cite{P52023} introduces a unified text-to-text framework for recommendation, converting all modalities—user behaviours, reviews, metadata—into prompt-formatted natural language. Built on an encoder–decoder LLM, it defines multiple RS tasks as conditional generation problems, enabling multi-task pretraining and zero-shot inference via customised prompts. This “Text-as-Interface” approach sets a foundational design for prompt-centric MRS. ViML~\cite{ViML} addresses trimodal alignment by generating music descriptions from structured tag data through prompt engineering and ICL. It integrates these summaries with video features and few-shot textual prompts, enabling text-guided music retrieval. The method highlights the use of prompting for semantic alignment across modalities. LLM-Rec~\cite{LLM_Rec} applies prompting strategies (e.g., attribute tagging and summarisation) to enhance raw item descriptions, using GPT-3 and LLaMA-2. These generated texts improve the RS model inputs while preserving transparency and interpretability.

Task-Specific Instruction Generation includes works such as LLAMA-E~\cite{LLAMA_E}, which generates instruction–response pairs from structured product data for domain adaptation, and GIRL~\cite{GIRL}, which creates synthetic job descriptions from CVs using supervised prompting. 

In few-shot tagging, we find that TagGPT~\cite{TagGPT} and PIE~\cite{PIE} employ prompt-based pipelines for attribute or tag extraction from multimodal cues in zero-/few-shot settings, while IntellectSeeker~\cite{IntellectSeeker} refines vague academic queries through prompted rewriting.

Together, these methods demonstrate how prompting can serve as a lightweight yet flexible bridge across diverse input types, supporting structured alignment, retrieval, generation, and task adaptation in LLM-based MRS.

\subsubsection{Adapted Multimodal Fusion}
\label{sec:fusion_adapted}
\paragraph{\textbf{Adapted Multimodal Fusion}}: These methods integrate multiple modalities—typically vision, text, and structured data—before passing inputs to an LLM, using pretrained encoders, adapters, or projection modules. They allow tighter cross-modal alignment, improved transferability, and richer representations, but often require additional components and training steps.

\textbf{PMMRec}~\cite{PMMRec} exemplifies full fusion without ID reliance. It encodes text and image features using multilingual RoBERTa and CLIP-ViT, respectively, then combines them via a Transformer-based fusion module to form item representations. These are consumed by a user encoder trained on contrastive and predictive objectives, producing embeddings directly usable by downstream LLMs. This design allows the model to convert heterogeneous item content into aligned latent representations suitable for recommendation without relying on discrete item identifiers. VIP5~\cite{VIP5} transforms multimodal inputs—including image embeddings, semantic IDs, and text—into prompt-compatible token sequences. It injects these into a frozen LLM backbone using adapters and soft prompt tokens, enabling direct, sequential, and explanatory recommendations. A learned projection module aligns visual features to the language space, ensuring efficient integration and modular personalisation. TMF~\cite{TripletFusion2024} fuses visual, textual, and graph signals through a multi-level attention architecture. Each modality is first encoded separately, then aligned through learned adapters. The resulting fused representation is inserted into the LLM input via specialised tokens, supporting instruction-tuned recommendation in multi-behaviour scenarios.

Other methods share similar goals with different fusion pipelines. MISSRec~\cite{MISSRec} uses modality-specific adapters to process frozen CLIP features and learns multimodal user sequences via contrastive objectives. UniMP~\cite{UniMP} fuses heterogeneous user inputs (e.g., text, images, metadata) using vision adapters and soft prompts for personalised generative tasks. MM-Rec~\cite{MM_Rec} applies co-attentional transformers on visual (ROI) and textual news data to build a shared user-item space.

Peripheral approaches such as HiRMed~\cite{HiRMed} (RAG over medical embeddings), Multi-LLM-GNSS~\cite{Multi_LLM_GNSS} (projected visual embeddings to LLM token space for an MLLM), and LLLM-Rec~\cite{LLLM_Rec} (textual embedding for retrieval) hint at potential extensions but do not perform explicit fusion.

Overall, these techniques offer robust multimodal alignment by injecting learned, fused embeddings into LLMs—bridging vision, language, and structured data for personalised and context-aware recommendation.

\paragraph{\textbf{Code-like Structural Conversion}}: This approach structures multimodal inputs—text, categorical fields, or images—into code-like formats such as JSON or Python-style objects before prompting. Though less explored, it offers a human-readable and programmatically parsable interface for LLM conditioning. \textbf{Mender}~\cite{Mender} exemplifies this strategy by converting structured and unstructured features into JSON-formatted prompts. Textual and categorical fields are serialised, while visual inputs are quantised into semantic IDs via RQ-VAE. These representations are fused in a cross-attentive TIGER-based architecture, supporting controlled, preference-aware generation. This emerging format may offer benefits in modularity, transparency, and schema alignment for LLM-based MRS.

\section{MRS methods}
\label{sec:mrsmethods}
This section focuses on strategies designed specifically for adapting LLMs to the needs of multimodal recommender systems (MRS). Unlike general-purpose LLM techniques, the methods discussed here address core challenges of MRS: combining heterogeneous modalities, preserving distinct modality semantics, and fusing user-item interactions into meaningful recommendations. We organize these approaches into three categories: Disentangle, Alignment, and Fusion.

Disentangling approaches aim to explicitly separate modality-specific from shared information, promoting better generalisation and interpretability. Alignment techniques focus on learning a shared representation space across modalities, typically using contrastive or encoder-based methods. Fusion strategies, on the other hand, determine when and how modality information is merged relative to the LLM, and at what granularity.

While these methods enhance MRS performance by leveraging modality-specific structure, they often come at the cost of architectural complexity, larger computational graphs, and additional pretraining or supervision. For instance, contrastive alignment and fine-grained fusion can require modality masking, negative sampling, or large batch sizes to be effective, which increases resource demands (e.g., MMGCL \cite{MMGCL2022}, P5 \cite{P52023}, MMRec \cite{MM_Rec}).

These MRS-specific strategies complement generic prompting or adaptation methods and are often key to unlocking high-quality multimodal recommendations.

These MRS-specific strategies complement generic prompting or adaptation methods and are often key to unlocking high-quality multimodal recommendations. While this survey avoids detailed architectural analysis, it is worth noting that recent MRS models increasingly adopt Graph Neural Networks and Transformer architectures to capture better complex user–item and modality interactions, as extensively studied in prior works~\cite{CARCA, SASRec, MBHET, linrec2023, BERT4Rec, AC_TSR, PositionalEncodingSRS}.

\subsection{Disentangle}
\label{sec:disentangle}
Disentangling is a crucial process in multimodal learning, aimed at separating shared and modality-specific information within the representations of different data types. The objective is to clearly distinguish features relevant across modalities from those unique to individual modalities. Various disentangling methods have been proposed, each pursuing this separation through different mechanisms. In this section, we analyse their application in multimodal systems, particularly in recommendation tasks where efficient integration of heterogeneous features is essential for performance and interpretability.

\begin{figure*}[t]
    \centering
    \includegraphics[width=0.85\textwidth]{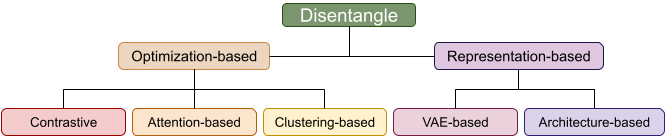}
    \caption{Disentangle categories. Use-intent is not considered a category; it is just a note at the end of the section.}
    \label{fig:disentangle_diagram}
\end{figure*}

Disentangling strategies can generally be divided into two types: \textbf{Representation-based disentangling}, which manipulates the feature space to separate common and specific components via dedicated loss functions, and \textbf{Optimisation-based disentangling}, which adds training constraints, such as orthogonality or alignment losses, to encourage independent component learning (Fig. \ref{fig:disentangle_diagram}). These strategies are often combined (e.g.,~\cite{PAMD2022}), and in this survey, we organise the reviewed works based on their architectural designs while indicating the disentangling mechanisms employed.

\subsubsection{Contrastive methods}
\label{sec:contrastive_disentangle}
\paragraph{\textbf{Contrastive methods}} These methods use contrastive learning to align or separate modality-specific and shared representations by leveraging positive/negative sample construction across modalities or views. This is one of the most common approaches since contrastive learning has been well known for some years, and these approaches are easily applicable to pretrained visual and text encoders. One common strategy is to manipulate different modalities to generate specialised negative samples.

NegGen~\cite{NegGen} proposes a causal learning framework for MRS that disentangles meaningful user preferences from spurious modality correlations through contrastive interventions. It generates negative samples by selectively altering key item attributes across text and visual modalities using MLLMs, followed by encoding and causal analysis to compute the causal effect of each change on recommendation scores. MMGCL~\cite{MMGCL2022} introduces contrastive modality disentanglement through techniques like modality masking and edge dropout. By perturbing specific modalities to create hard negative examples and randomly dropping inter-modality connections during graph-based learning, MMGCL encourages the model to learn independent features for each modality while preserving their shared representations. Taobao~\cite{TaobaoMRS_2024} tackles modality disentanglement in ID-based systems by introducing Semantic-Aware Contrastive Learning (SCL). SCL optimises orthogonality and alignment between modality-specific representations by emphasising semantic similarity across visual and textual modalities.

In addition to these methods, several other approaches apply contrastive techniques to promote disentanglement. MICRO~\cite{MICRO2022} constructs separate modality-aware item graphs and uses contrastive learning to align each modality’s representation with a fused multimodal representation, preserving both shared and specific traits. PMMRec~\cite{PMMRec} develops a contrastive framework (NICL) that enhances intra- and inter-modality alignment without relying on ID features, contributing to robust recommendation; it will be further discussed in Section~\ref{sec:alignment}.

Finally, although not strictly multimodal, CrossCBR~\cite{CrossCBR2022} proposes a disentanglement method between user and bundle representations using separate interaction graphs and cross-view contrastive learning. While originally focused on bundles and items, the approach could be extended to multimodal relations by constructing modality-aware graphs.

\subsubsection{Attention-Based methods}
\label{sec:attention_disentangle}

\paragraph{\textbf{Attention-Based methods}}
These approaches apply attention mechanisms to reweight or isolate modality contributions, often in conjunction with knowledge distillation or causal analysis, to disentangle signal relevance dynamically. One of the first examples is MAML~\cite{MAML2019}, which applies attention to adaptively assign importance to textual and visual modalities based on user preferences.

PromptMM~\cite{promptmm} proposes a modality-aware list-wise knowledge distillation framework that dynamically reweights modality contributions during ranking, enabling fine-grained disentanglement of collaborative and multimodal signals. It transfers structured preference knowledge from a complex teacher to a lightweight student via embedding, pairwise, and listwise distillation, using an attention-like reweighting formulation. RecInDial~\cite{RecInDial} disentangles entity-based user preferences from conversational context by leveraging a knowledge graph encoded with R-GCN and an entity-level attention mechanism. The entity-based representation is injected as a bias during dialogue generation, with a pointer mechanism maintaining flexible separation between entities and conversation flow. MMContrastive~\cite{MMContrastive2021} learns separate item representations for text, image, and graph modalities, aligning text and image via a contrastive loss and fusing them with additive attention. The model preserves modality-specific features while learning shared semantics, facilitating disentangled multimodal understanding.

\subsubsection{Clustering-Based methods}
\label{sec:cluster_disentangle}

\paragraph{\textbf{Clustering-Based Disentanglement}}
These models use clustering mechanisms on tokens, items, or users to capture shared and divergent interests across modalities or subgroups, often enhancing personalisation or structure learning. This approach was among the earliest for modality disentanglement and alignment, as seen in early methods like LATTICE.

MISSRec~\cite{MISSRec} introduces a multi-modal interest discovery (MID) module that applies density-based clustering over item tokens to generate interpretable interest prototypes reflecting diverse user preferences. These prototypes are incorporated into a Transformer-based encoder-decoder framework, enabling sequence-level modelling of modality-specific and cross-modality interests while improving personalisation through disentangled interest representations. HCGCN~\cite{HCGCN2022} proposes a hybrid disentanglement strategy by separating visual and textual item features into dual clustering-based item-item graphs and aligning them via a multimodal contrastive objective. It further models user-item preferences through both local co-cluster GCNs and a global hierarchical user-item graph, promoting semantic alignment and diversified user behaviour modelling across clusters.

\subsubsection{VAE-Based methods}
\label{sec:vae_disentangle}

\paragraph{\textbf{VAE-Based methods}}
These methods use (variational) autoencoders to learn latent factors, enforcing structural independence between content, collaborative, and modality-specific representations. In multimodal systems, VAE-based approaches play a critical role in disentangling underlying structures, either at the latent factor level across modalities or between content and collaborative signals.

\paragraph{\textbf{Latent Factor Disentanglement \& Interpretability}}
Some multimodal systems perform disentanglement at the latent factor level, aiming to separate semantically meaningful components rather than directly separating modalities. DGVAE~\cite{DGVAE} enhances recommendation interpretability by disentangling latent user-item interactions into prototype-based factors (e.g., brand, category) using graph variational autoencoders. It leverages mutual information maximisation to align factors derived from ratings and text content, while a frozen multimodal item-item graph and residual GCNs help structure this factorisation.

MVGAE~\cite{MVGAE2021} models modality-specific uncertainty via independent variational encoders for each modality, outputting Gaussian distributions where variance reflects confidence. A product-of-experts mechanism fuses modality posteriors, weighting more reliable modalities more heavily, followed by a second fusion step that incorporates collaborative information into uncertainty-aware multimodal representations. While not a VAE-based method, SampleLLM~\cite{SampleLLM} introduces a two-stage disentanglement process for synthetic tabular data generation, using feature attribution to isolate interacting feature groups. Although operating within the tabular domain, this work shares the goal of semantic factor separation, aligning real and synthetic distributions by identifying meaningful dependencies.

\paragraph{\textbf{Content-Collaborative Disentanglement}}
Other VAE-based approaches focus on separating content-driven and collaborative signals, offering robustness and potential extensions to multimodal scenarios. DICER~\cite{DICER2020} presents an early framework that enforces orthogonality between content-derived and interaction-derived features, while also promoting independence across latent dimensions using KL-based regularisation. Although not multimodal, this structure laid the groundwork for later extensions.

At a more structured level, MacridVAE~\cite{MacridVAE} achieves hierarchical disentanglement by dividing user intent into macro-concepts (e.g., categories) and fine-grained preferences (e.g., style), using a combination of prototype clustering and $\beta$-VAE regularisation. This architecture allows controllable manipulation of latent factors during recommendation. An advanced evolution of hierarchical VAE design is found in Mender~\cite{Mender}, which integrates LLM-based language conditioning with discrete semantic representations through an RQ-VAE architecture. Mender explicitly disentangles user preferences from historical interactions, capturing modality-specific item features as hierarchical latent codes and enabling fine-grained, interpretable control over sequential recommendation generation.

\subsubsection{Architecture-Driven methods}
\label{sec:architecture_disentangle}
Methods that build architectural components, such as specialised encoders, routing modules, or mixture-of-experts setups,to separate and control modality- or intent-specific signals.

\paragraph{\textbf{Modality-specific Encoders and Decomposition}}
One prominent approach is designing dedicated encoders or decomposition modules to split modality-common and modality-specific representations. PAD~\cite{PAD2024} introduces a three-phase framework for multimodal sequential recommendation, culminating in a triple-expert Mixture-of-Experts architecture. After separately pretraining a collaborative (ID-based) expert and a semantic (LLM2Vec-based) text encoder, PAD aligns them using a multi-kernel MMD loss and fine-tunes three dedicated experts—collaborative-specific, text-specific, and aligned—through an adaptive gating mechanism. This structure explicitly preserves modality-specific traits while enabling flexible integration based on item frequency.

A related but earlier approach is PAMD~\cite{PAMD2022}, which decomposes item modalities into common and specific components using pretrained text and image encoders. Commonality is encouraged through direct alignment losses, while modality-specific features are enforced to be orthogonal. This two-step disentanglement mechanism is reinforced with an auxiliary contrastive loss to better preserve shared and independent characteristics. Other works apply separate encoders without a full decomposition step. ViML~\cite{ViML} independently encodes video, audio, and text modalities before fusion, preserving distinct semantic information across modalities. A text dropout mechanism is additionally used to prevent over-reliance on text during retrieval tasks, though the disentanglement strategy is more straightforward than specialised decomposition architectures.

\paragraph{\textbf{User Intent Disentanglement}}
While not directly focused on multimodal inputs, some approaches disentangle different aspects of user motivation through feedback-aware modelling. IntentDisentangle~\cite{IntentDisentangle} separates popularity-driven and preference-driven user behaviours using intent prototypes and clustering over collaborative and content features, combined with VAE-based regularisation. Similarly, CDR~\cite{CDR2021} applies separate Transformer encoders to clicked, unclicked, and disliked feedback, leveraging co-filtering and curriculum learning to form disentangled user intention vectors.

\subsection{Alignment}
\label{sec:alignment}
Alignment strategies play a central role in multimodal learning by enforcing coherence between heterogeneous modality representations by projecting them into shared semantic spaces or coordinating their interactions during model inference or training. The ultimate goal is to ensure that modality-specific representations are aligned to capture shared semantics and enable consistent cross-modal reasoning. This is particularly important in recommendation systems, where aligned embeddings facilitate joint modelling of user preferences, item content, and interaction behaviours across different data types.

Approaches to alignment vary significantly in terms of mechanism, flexibility, and computational cost. Some strategies operate at the optimisation level by aligning representations through losses such as contrastive or reconstruction-based objectives. Others rely on architectural additions—such as adapters, projection heads, or modular encoders—to transform or bridge modalities. More recently, instruction tuning, prompting, and reinforcement learning have been explored as lightweight alternatives to full model retraining, offering scalable alignment across diverse data types with minimal modification to large language models (LLMs).

\begin{figure*}[t]
    \centering
    \includegraphics[width=0.92\textwidth]{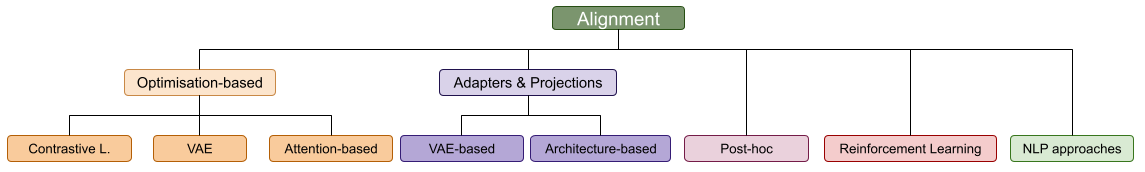}
    \caption{Alignment strategies.}
    \label{fig:alignment_diagram}
\end{figure*}

We organize the alignment methods surveyed in this section into five categories based on the mechanism and architectural design, which can be found in the Figure \ref{fig:alignment_diagram}. These categories are not mutually exclusive—certain models span multiple alignment mechanisms. Moreover, there is intentional overlap with Section~\ref{sec:disentangle}, where some alignment methods focused primarily on separating shared and modality-specific signals are discussed. To reduce redundancy, several such papers are only briefly mentioned here.

Finally, alignment methods vary in cost and applicability. Optimisation based and post-hoc approaches often require access to full training loops and modality-specific encoders, making them more resource-intensive. In contrast, adapter-based and NLP-aligned techniques are lighter and more modular, facilitating scalable deployment in large-scale or real-time recommendation systems without full retraining.

\subsubsection{Optimisation-based Alignment}
Optimisation-based alignment includes techniques that optimize specific objectives, such as contrastive learning, cross-modal attention, or variational inference, to bring modality representations closer.

\paragraph{\textbf{Contrastive Learning Alignment}}
In several recent works, contrastive learning (CL) serves as a core mechanism to align multimodal representations. HLLM~\cite{HLLM2024} represents a strong example, enhancing multimodal sequential recommendation by fine-tuning a Multimodal LLM (MLLM) with contrastive learning based on negative sampling. By distinguishing relevant and irrelevant user-item pairs during supervised fine-tuning, HLLM enables precise user preference modelling across modalities. Similarly, RLMRec~\cite{RLMRec2024} combines contrastive alignment (InfoNCE loss) with masked reconstruction to maximise mutual information between collaborative and LLM-derived semantic embeddings, effectively injecting structured semantic knowledge into collaborative representations.

PMMRec further explores contrastive alignment by jointly modelling inter-modality and intra-modality negative samples between text and image features, while incorporating next-item positives to capture sequential transitions in recommendation patterns better. These strategies highlight different ways of embedding semantic coherence and behaviour patterns into multimodal representations through contrastive mechanisms.

Other approaches, such as K-RagRec~\cite{KRagRec} and CTRL~\cite{CTRL2023}, also employ contrastive-like alignment strategies, though they are either less strictly multimodal or focus more on structural adaptation via external graphs or tabular data transformation.

Several additional works leveraging contrastive learning for multimodal alignment have already been discussed in the Disentangle section (Section~\ref{sec:disentangle}). These include ViML~\cite{ViML}, MMContrastive~\cite{MMContrastive2021}, and SampleLLM~\cite{SampleLLM}, which apply contrastive or contrastive-like objectives to align multiple modalities or regularize synthetic data generation. Furthermore, models such as HCGCN~\cite{HCGCN2022}, NegGen~\cite{NegGen}, MICRO~\cite{MICRO2022}, MMGCL~\cite{MMGCL2022}, and Taobao~\cite{TaobaoMRS_2024} integrate contrastive strategies primarily aimed at disentangled or modality-invariant representation learning.

\paragraph{\textbf{VAE Aligment}}
Variational Autoencoder (VAE)-based strategies offer a powerful mechanism for aligning multimodal representations by modelling latent distributions across modalities.
TTDS~\cite{TTDS2024} exemplifies this approach by proposing a dynamic semantic indexing framework for LLM-based recommendation. It generates semantic token sequences for users and items via a twin-tower token generator and aligns them with collaborative signals using a dual-modality VAE, enabling high-order user-item pattern modelling and end-to-end fine-tuning.
Related VAE-based models such as MVGAE~\cite{MVGAE2021} and DGVAE~\cite{DGVAE} have also been discussed in Section~\ref{sec:disentangle}, emphasizing disentangled or distributionally aligned representations across modalities.

\paragraph{\textbf{Attention-based alignment}} techniques leverage cross-modal interactions to fuse heterogeneous modalities in a structured and interpretable manner. UniMP~\cite{UniMP} integrates multimodal user histories by applying cross-attention between visual and textual embeddings, complemented by multi-modal soft prompts to enhance personalisation in sequential modelling tasks. Similarly, MM-Rec~\cite{MM_Rec} adopts a ViLBERT-style co-attentional transformer architecture to align and fuse textual and visual news representations, enabling unified multimodal embeddings optimised for downstream recommendation tasks.


\subsubsection{Adapter and Projection-based Alignment}
Adapter and Projection based Alignment introduces lightweight neural components (e.g., adapters or linear projection layers) that flexibly map modality-specific inputs into shared latent spaces.

\paragraph{\textbf{Adapter Alignment}}
Adapter-based techniques provide flexible mechanisms to align multimodal information by introducing specialised intermediate modules without fully retraining large models. Among these, iLoRA~\cite{iLoRA} stands out by incorporating a dynamic gating mechanism that customises expert selection based on each user's interaction sequence. This enables fine-grained alignment to individual user preferences, improving adaptability and mitigating negative transfer across diverse behaviour patterns.

KAR~\cite{KAR} also leverages adapters, proposing a framework where external factual knowledge extracted from an LLM is transformed through a hybrid expert adaptor into dense representations suitable for recommendation models. This bridging of LLM outputs and recommender systems highlights how adapters can flexibly map semantic information into structured predictive spaces. Additionally, ADKGD~\cite{ADKGD} explores adapter-like designs by enforcing disentangled representations of knowledge graph triplets via dual-view Bi-LSTM encoders and a VAE-style alignment loss, though its primary focus is on anomaly detection rather than recommendation.

\paragraph{\textbf{Semantic or Projection Alignment}}
Projection-based and semantic alignment methods aim to map diverse modalities into coherent shared spaces, facilitating downstream multimodal reasoning.
Among these, Multi-LLM-GNSS~\cite{Multi_LLM_GNSS} is a representative example, aligning CLIP-derived visual embeddings with LLM token embeddings via a learnable projection layer, followed by fine-tuning the LLaVA model jointly on visual and textual data. This approach enables robust multimodal reasoning over GNSS snapshots, addressing interference monitoring and classification challenges.

PROMO~\cite{PROMO} also adopts a projection-based design but focuses on the cold-start problem in recommendation, projecting diverse modalities such as ID embeddings, numerical features, and user feedback into a unified latent space through its Personalised Prompt Network (PPN). This lightweight, independent encoder design enhances adaptability across item types. In contrast, IntellectSeeker~\cite{IntellectSeeker} applies semantic alignment techniques by structuring user queries and academic article features into a common semantic space. While effective for retrieval tasks, it primarily emphasises textual semantic enhancement rather than true multimodal integration.


\subsubsection{Post-hoc Alignment}
\paragraph{\textbf{Post-hoc Alignment}}
Post-hoc alignment strategies refine representations after initial modelling, ensuring better consistency between multimodal or semantic spaces.
Molar~\cite{Molar} proposes a decoupled architecture where multimodal item modelling and user embedding generation are performed independently, followed by a contrastive alignment step between ID-based and content-based user embeddings. This design preserves modality-specific signals during early stages and only aligns them at the final stage, effectively disentangling collaborative and semantic information. In a related line, LGIR~\cite{LGIR} employs a GAN-based refinement mechanism, where a generator learns to transform low-quality LLM-derived resume embeddings into higher-quality ones, and a discriminator enforces the alignment through adversarial training. Although developed for resume recommendation, this adversarial post-hoc refinement approach shares similar motivations of enhancing representation quality without retraining the full system.


\subsubsection{RL-Enhanced and External Alignment}
\paragraph{\textbf{RL-Enhanced and External Alignment}}
Incorporating reinforcement learning and external reasoning mechanisms provides additional pathways for aligning multimodal or heterogeneous information with user intent.
GIRL~\cite{GIRL} applies reinforcement learning by training a reward model to score CV–job description matches, followed by fine-tuning an LLM generator using Proximal Policy Optimisation (PPO) based on these rewards. AgentRec~\cite{AgentRec} similarly leverages Reinforcement Learning from Human Feedback (RLHF) to fine-tune an SBERT encoder, aligning sentence embeddings for prompts intended for the same agent and enhancing agent retrieval performance.


\subsubsection{NLP-based Alignment}
\paragraph{\textbf{NLP-based Alignment}}
Fine-tuning and instruction tuning enable LLMs to adapt to recommendation tasks by aligning user, item, and modality-specific features within the model's internal representations.
HKFR~\cite{HKFR} and LLaMA-E~\cite{LLAMA_E} both apply instruction tuning approaches: HKFR focuses on heterogeneous user behaviour logs, transforming structured data into textual inputs to enhance recommendation reasoning, while LLaMA-E designs specialised prompts that interleave seller, customer, and platform features for content generation and classification tasks in e-commerce.

Going beyond textual alignment, VIP5~\cite{VIP5} maps image features into token embeddings fused with item text tokens through adapter tuning, achieving implicit multimodal alignment without modifying the backbone LLM. Similarly, TMF~\cite{TripletFusion2024} integrates visual, textual, and graph modalities through a two-stage attention mechanism, producing a unified embedding space optimised for multi-behaviour recommendation tasks. Complementarily, IndustryScope~\cite{IndustryScopeGPT2024} introduces a tool-assisted reasoning framework, where an LLM interacts with external multimodal knowledge graphs through structured tool outputs, achieving indirect alignment without modifying the LLM itself.

\subsection{Fusion Classification}
\label{sec:fusion}

In multimodal recommendation systems (MRS), the vast and diverse multimodal information associated with users and items necessitates effective fusion techniques to generate comprehensive feature vectors for recommendation tasks~\cite{IMF2023}. Historically, fusion methods have evolved from concatenation and simple networks to join the modalities to attention mechanisms, and these from coarse-grained to fine-grained approaches, each addressing different aspects of multimodal data integration. This evolution was a key focus in previous surveys on MRS (\cite{MRS1_survey, MRS2_survey}), highlighting increased diversity among publications. Early classifications of fusion approaches considered three main features: the \textbf{location of fusion}, the \textbf{graphs involved} 
, and the \textbf{type of fusion}. In this section, we will review these categories, except the one referring to graphs, because of their low relevance for this survey, refreshing the previous work for MRS that can be useful in the future. We will also dive a little deeper into what the previous surveys did in the attention mechanisms to understand better how they have been used and to be able to extract these ideas for the future.

\subsubsection{Location of Fusion}
In earlier surveys where Large Language Models (LLMs) were not the primary component, the \textbf{location of fusion}---\textbf{early}, \textbf{intermediate}, and \textbf{late}---was a crucial classification criterion.
\paragraph{\textbf{Location of LLMs}} 
However, with the advent of LLM-based approaches, fusion occurs differently. When considering the LLM as the central component, most approaches integrate multimodal information \textbf{before} the LLM or \textbf{in parallel} with it. For instance, some methods convert modalities into text representations before feeding them into the LLM~\cite{TripletFusion2024, HLLM2024, P52023}, while others facilitate interactions between agents and knowledge graphs (KG) as in IndustryScopeGPT~\cite{IndustryScopeGPT2024}. 
Additionally, techniques like soft prompting, used in TripletFusion~\cite{TripletFusion2024}, are employed to integrate multimodal data. Conversely, parallel methods involve \textbf{alignment} mechanisms, as seen in Section \ref{sec:alignment} for CTRL~\cite{CTRL2023} and PAD~\cite{PAD2024}, which rely on Contrastive Learning (CL) and Variational Autoencoders (VAEs), or through FT strategies (cites).

Still, there are new models that use LLMs but use intermediate fusion. ViML~\cite{ViML} employs a Transformer-based architecture to seamlessly integrate video and text representations, enabling effective querying of music samples by combining visual and linguistic information. But the roll of the LLMs in this case is Data Augmentation.

\paragraph{\textbf{Early Fusion}} refers to the integration of multiple modalities at the data level, typically by concatenating or summing modality-specific features before inputting them into the model (\cite{VBPR2016,UVCAN2019,MAML2019,LATTICE2021,FREEDOM2023,HCGCN2022}). These techniques aim to capture shared patterns across modalities, enhancing the robustness of the unified representation, but overlook complementary modality-specific information.

\paragraph{\textbf{Intermediate Fusion}} involves integrating modalities after feature extraction but before the final decision-making layer, often leveraging attention mechanisms to weight the importance of each modality dynamically. This category includes concatenation-based methods like MMGCL~\cite{MMGCL2022}, attention-based approaches such as MMRec~\cite{MMRecSM2023}, MKGAT~\cite{MKGAT2020}, and DMRL~\cite{DMRL2022}, as well as Product of Expert models like MVGAE~\cite{MVGAE2021} and Ego Fusion methods like EgoGCN~\cite{EgoGCN2022}. These methods strike a balance between early and late fusion by allowing interactions between modalities at a more granular level within the model architecture.

\paragraph{\textbf{Late Fusion}} combines modality-specific predictions or representations at the decision level, maintaining modality independence during processing. This can be achieved through concatenation \cite{MMGCN2021, GRCN2020, SLMRec2022}, element-wise sum \cite{MMGCN2021, SLMRec2022}, and attention-based methods  \cite{Dualgnn2021,MGAT2020,PAMD2022,DRAGON2023}. Late fusion leverages the strengths of each modality independently, ensuring that the fusion process does not introduce errors from one modality affecting another.


\subsubsection{Type of Fusion}
Another pivotal classification feature in MRS fusion methods is the \textbf{type of fusion}, which encompasses the architectural and methodological aspects of integrating multimodal data. Initially, many approaches utilised various join methods like concatenation (\cite{MMGCN2021,MAML2019,VBPR2016,MMGCN2021}), or combination of experts (\cite{MVGAE2021}), but later they were mainly attetion-based mechanisms (\cite{MGAT2020, MKGAT2020, HCGCN2022, UVCAN2019, CMBF2021, POG2019, NOR2019, LATTICE2021, Dualgnn2021, MGAT2020, PAMD2022, DRAGON2023}). Early models implemented the \textbf{Bahdanau Attention} mechanism \cite{Bahdanau2014}, which is \textbf{additive} in nature, allowing for flexible alignment by combining query and key vectors through a non-linear function. Subsequently, there was a shift towards \textbf{multiplicative attention} mechanisms \cite{Luong2015}, which utilise dot products for computational efficiency.

As the field progressed, the limitations of \textbf{coarse-grained fusion}---such as the potential loss of modality-specific features and the invasive nature of merging disparate data---prompted the development of \textbf{fine-grained fusion} techniques. Fine-grained fusion typically employs a combination of different attention mechanisms, including \textbf{cross-attention} and \textbf{co-attention}, to preserve and integrate specific features from each modality \cite{MGAT2020, MKGAT2020}. This approach is essential for maintaining the integrity of unique modality characteristics, particularly in domains like fashion recommendation, where detailed features such as clothing patterns or audio tones are critical \cite{POG2019, NOR2019}. As a complementary approach, some papers adapted self-attention, making it less invasive \cite{NOVA2021, MML2022} or its combination with the cross and co-attention \cite{POG2019,CMBF2021, NOR2019}

\textbf{Coarse-Grained Fusion} represents the early stage of fusion methods, primarily focusing on the common features across different modalities. These approaches typically merge multimodal information at a higher, more aggregated level, emphasising shared characteristics to create unified representations. For example, UVCAN~\cite{UVCAN2019} segregates multimodal data into user-side and item-side information, employing additive co-attention to generate fusion weights that emphasise commonalities between these sides. In contrast, CMBF~\cite{CMBF2021} combines self-attention with cross-attention by splitting data based on modality rather than user-item divisions. The self-attention mechanism first extracts intra-modality information, which is then combined with cross-attention to facilitate inter-modality interactions. MML~\cite{MML2022} is a feature-aware self-attention approach that extends traditional self-attention by integrating multimodal side information through an additive attentional bias applied to the user and item IDs.

As the limitations of coarse-grained fusion became apparent, \textbf{fine-grained fusion} techniques were developed to preserve the unique characteristics of each modality by selectively integrating detailed, modality-specific information. This approach is particularly significant in domains like fashion recommendation, where nuanced features such as clothing patterns or audio tones are critical. For instance, POG~\cite{POG2019} trains a classic transformer-based model with self-attention in the encoder and cross-attention in the decoder using masking techniques to achieve fine-grained integration of fashion image features. NOR~\cite{NOR2019} enhances this by using an encoder-decoder transformer with fine-grained self-attention and mutual attention (co-attention) instead of standard cross-attention, facilitating the generation of descriptive schemes based on collocation information. Other models like EFRN~\cite{EFRM2019} utilize Semantic Extraction Networks (SEN) to extract local features, and MGAT~\cite{MGAT2020} applies gated attention mechanisms to enhance intra-modality features. For LLMs we have iLoRA~\cite{iLoRA},~\cite{TripletFusion2024}.

Fine-grained fusion often involves \textbf{cross-attention} and \textbf{co-attention} mechanisms to preserve modality-specific features. Cross-attention concentrates the attention of one sequence on another, facilitating the preservation of specific modality features. Co-attention, on the other hand, enables bidirectional attention by combining two cross-attention modules, allowing both sequences to attend to each other simultaneously \cite{coattention2016}. This bidirectional approach ensures a more integrated and comprehensive understanding of the multimodal data.

In a middle point between coarse and fine-grained, we find models that combine both, as CMBF~\cite{CMBF2021}, thereby preserving global aggregation while maintaining detailed modality-specific features. For example, NOVA~\cite{NOVA2021} introduces a non-invasive attention mechanism with dual branches (V for the IDs and Q and K for modalities) to retain interactive information during fusion, addressing the shortcomings of vanilla attention approaches. Similarly, MMContrastive~\cite{MMContrastive2021} implements a simple average pooling for intra-modality and additive cross-attention to address inter-modality features.
Additionally, some models incorporate \textbf{gated attention} mechanisms to control the flow of information from different modalities dynamically. For instance, MGAT~\cite{MGAT2020} utilizes a gated attention mechanism to focus on the user’s local preferences from \GUI, allowing the model to weigh the importance of each modality based on contextual relevance. 

Overall, the evolution of fusion methods in MRS highlights a transition from \textbf{coarse-grained} approaches that emphasise commonalities across modalities to \textbf{fine-grained} techniques that preserve and integrate specific features of each modality. This progression underscores the importance of balancing global aggregation with detailed, modality-specific information to achieve robust and accurate recommendations.


\section{Main Trends and Future Directions}
\label{sec:future}

As the field of multimodal recommendation systems (MRS) continues to evolve, several key trends are shaping the future of model integration and representation learning. These emerging directions focus on optimizing multimodal interactions, improving efficiency, and leveraging more sophisticated techniques for data fusion, prompting, and model adaptation.

\begin{enumerate}
    \item \textbf{KGs as a Bridge for Multimodal Integration} \\
    Knowledge Graphs (KGs) provide a structured intermediary for aligning heterogeneous modalities—including text, images, tabular data, or even geospatial information—into a unified representation. Recent approaches like IndustryScopeGPT \cite{IndustryScopeGPT2024} and K-RagRec~\cite{KRagRec} demonstrate how converting diverse inputs into KG triplets enables effective grounding, retrieval, and reasoning with LLMs. Future work should explore scalable KG construction pipelines, dynamic subgraph retrieval, and fine-grained alignment between KG representations and LLM embeddings. This direction may enhance generalisation, support real-world context, and reduce hallucination in multimodal recommendation.

    \item \textbf{Combination of soft prompting techniques and Adapters (LoRA, QLoRA) as a standard fusion method} \
    Combining soft prompting with adapter-based fine-tuning techniques like LoRA and QLoRA offers a promising direction for modular and efficient multimodal fusion. These methods enable targeted, low-rank adaptation of LLMs, facilitating the integration of modality-specific representations without updating the entire model. This especially appeals when parameter efficiency, continual learning, or privacy constraints are essential. For instance, \cite{iLoRA} applies individualised LoRA adapters to user representations, and this idea could be naturally extended to handle different modalities such as vision or structured knowledge. Similarly, soft prompts—whether semantic ID vectors \cite{PersonalizedPromptLearning} or learnable embeddings—can encode user- or modality-specific context. Recent works like \cite{TripletFusion2024} suggest that combining different forms of lightweight tuning (e.g., soft prompts for users and adapters for modalities) improves generalisation across cold-start and unseen input configurations. Despite its promise, this strategy introduces new questions around prompt–adapter interaction, fusion order, and tuning interference, warranting further theoretical and empirical investigation.
    
    \item \textbf{Extension of the usage of attention mechanisms combining coarse and fine-grained attention based on critical topics} \\
    Future research may explore combining coarse and fine-grained attention mechanisms in a more dynamic way, adapting them to the specific needs of each multimodal task. By focusing attention on critical topics, models can more effectively process and integrate diverse inputs, improving both relevance and accuracy in recommendations.

    \item \textbf{Masking is not outdated, just renewed} \
    Masking techniques like MLM remain valuable in LLM-based recommendation, especially when adapted for multimodal data. Recent work such as \cite{MMGCL2022} shows that masked contrastive learning can effectively align modalities. Combining masking with fine-tuning or soft prompting offers greater flexibility and control, particularly in low-resource or unsupervised settings. While training can be more costly than prompt-based methods, masking remains a robust and adaptable tool for multimodal fusion.
    
    \item \textbf{LLMs for alignment through summarisation} \\
    Several works like \citet{HLLM2024} and \citet{Molar} use image-to-text summarisation to align modalities with LLMs. Future research could explore LLMs as semantic summarisers for modality alignment, offering an interpretable, encoder-free alternative to contrastive learning or projection layers, inspired by early ideas in \citet{CTRL2023}.

    \item \textbf{MLLMs for Recommendation} \\
    Many works, such as VIP5~\cite{VIP5}, UniMP~\cite{UniMP}, and Molar~\cite{Molar}, have shown the promise of MLLMs in modeling diverse modalities through fusion mechanisms or end-to-end pipelines. However, most of these systems operate in a frozen or lightly fine-tuned state and remain impractical for real-time recommendation due to their inference overhead. As a result, future work should explore hybrid architectures that use MLLMs for offline pretraining or content enrichment, while relying on lightweight alignment mechanisms (e.g., PMMRec~\cite{PMMRec}, MISSRec~\cite{MISSRec}) or modular fusion layers for deployment-time efficiency. Additionally, research into distilled MLLMs, modality-specific compression, and adaptive prompting could enable MLLMs to play a more practical role in production pipelines. There is also room to understand better the trade-offs between generalization, adaptability, and efficiency, especially in sparse or imbalanced modalities.
     
    \item \textbf{LLM-based Evaluation} \\
    While human evaluation remains a gold standard for assessing subjective qualities like helpfulness, coherence, and user satisfaction, it is costly and challenging to scale. Recent works such as GIRL \cite{GIRL} demonstrate using LLMs as evaluators, showing promise in approximating human judgment on natural language outputs. Future research can further explore LLM-based evaluation as a low-cost alternative, particularly for recommendation tasks with textual outputs (e.g., explanations, item summaries, dialogue). This direction invites systematic studies on the reliability, bias, and calibration of LLM evaluators compared to human judgments. Moreover, using LLMs enables new evaluation protocols, such as interactive or preference-based tests, and could be extended to multimodal setups, where LLMs are prompted with textual descriptions of images or videos for cross-modal assessments. Despite their efficiency, LLM evaluators introduce risks of alignment gaps and reproducibility issues, requiring careful benchmarking and transparency to ensure validity.

    \item \textbf{Structured JSON inputs for LLM reasoning} \\
    Recent works such as \cite{LLMsontheFly, leveragingllmreasoning, RLMRec2024} explore the use of structured JSON formats to introduce diverse data types (e.g., user profiles, item attributes, interaction logs) into LLM-based recommender systems. This structured input format helps LLMs better reason over relationships and constraints by making data semantics explicit. JSON inputs also enhance interoperability with APIs and support flexible prompt design. However, these methods require careful schema design, prompt formulation, and output validation to ensure the LLM handles the structure correctly and consistently.

    \item \textbf{Python class-based prompts for structured reasoning} \\
    Some recent NLP studies propose representing schemas and structured inputs using Python class definitions in the prompt \cite{KnowCoder, CODE4STRUCT, LLMIE}. This approach leverages LLMs' familiarity with programming constructs to encode complex relationships, type constraints, or inheritance patterns in a format that improves reasoning. These structured prompts are particularly useful for tasks like information extraction and event prediction. While they offer improved interpretability and modularity, their effectiveness depends on the LLM’s code understanding capabilities and require careful prompt engineering and validation.

    \item \textbf{Agents and RAG} \\
    Using agents in multimodal recommendation systems, particularly through retrieval-augmented generation (RAG), presents an exciting and underexplored future direction. These systems can autonomously query external databases or knowledge sources, enhancing recommendations by combining real-time retrieval with structured reasoning. For example, IndustryScopeGPT~\cite{IndustryScopeGPT2024} employs a Monte Carlo Tree Search (MCTS) framework with recursive prompting over a structured KG to simulate multi-step planning and decision-making. HiRMed~\cite{HiRMed} modularly integrates semantic retrieval (via FAISS), black-box LLM reasoning, and a fine-tuned model for recommendation prioritization, illustrating an agent-style orchestration pipeline. 
    
    Extending these methods toward persistent agent systems capable of interleaving retrieval, planning, and reflection, with explicit control over tools and modalities (e.g., images, structured data, KGs) could enhance interpretability, user alignment, and robustness. This approach also opens avenues for multi-agent simulation (e.g., RecSysLLMsP~\cite{RecSysLLMsP}) or agent selection strategies (e.g., AgentRec~\cite{AgentRec}), enabling dynamic behaviour modelling and task delegation. However, building such systems remains challenging due to the high inference cost, difficulty evaluating agent reasoning, and the lack of standardised datasets for agent-based recommendation. Despite these limitations, integrating RAG and agentic planning mechanisms remains a promising and novel research direction.
\end{enumerate}

\bibliography{custom}
\bibliographystyle{ACM-Reference-Format}

\clearpage
\onecolumn

\appendix
\section{Appendix}

\subsection{Datasets}
\label{sec:datasets}
Tables~\ref{tab:datasets1} and~\ref{tab:datasets2} present an overview of publicly available multimodal datasets used in multimodal recommender systems research.
Each dataset is categorised by its name, access link, modalities, domain and notes. We also include some other datasets used in the related papers. 

\begin{table*}[ht]
\centering
\small
\resizebox{\textwidth}{!}{%
  \begin{tabularx}{\textwidth}{l l l l X}
\toprule
Name & Link & Modalities & Domain & Notes \\
\midrule
Cora & \href{https://linqs.org/datasets/\#cora}{Link} & C, T & Academic & 2,708 ML papers in 7 classes with 5,429 citation links forming a graph (\cite{cora}) \\
Pubmed & \href{https://pubmed.ncbi.nlm.nih.gov/download}{Link} & C, T & Academic & Metadata from biomedical literature; requires preprocessing \\
Book-Crossings & \href{https://www.kaggle.com/code/jirakst/book-recommendation?utm_source=chatgpt.com}{kaggle}, \href{https://www.kaggle.com/datasets/jirakst/bookcrossing/data?select=BX-Books.csv}{kaggle}, \href{https://paperswithcode.com/sota/recommendation-systems-on-book-crossing-1}{paperswithcode}  & C, I, T & Book & \href{http://www2.informatik.uni-freiburg.de/cziegler/BX/}{old link} \\
MAVE & \href{https://github.com/google-research-datasets/MAVE}{github} & C, T & E-commerce & Includes real-world search queries and instructions (\cite{mave}) \\
Amazon Various & \href{https://cseweb.ucsd.edu/jmcauley/datasets.html\#amazon_reviews}{Link} & C, I, N, T & E-commerce & There exist many versions of these datasets. Some of them with images and text already encoded \\
YT8M-MusicTextClips & \href{https://zenodo.org/records/8040754}{Link} & I, T & E-commerce & Covers 7 product categories from Alibaba; includes item text, images, user interaction logs  \\
AliShop-7C & \href{https://jianxinma.github.io/disentangle-recsys.html}{github} & C, I, T & E-commerce & ~ \\
HM & \href{https://www.kaggle.com/competitions/h-and-m-personalized-fashion-recommendations}{kaggle} & I, T & E-commerce & ~ \\
bodyFashion  & \href{https://dxresearch.wixsite.com/pcw-dc}{Link} & I, T & Fashion & \cite{personalizedcapsule} \\
UTZappos50K  & \href{https://vision.cs.utexas.edu/projects/finegrained/utzap50k/}{Link} & I, T & Fashion & \citet{semanticjitter,finegrained} \\
iFashion (POG) & \href{https://github.com/wenyuer/POG}{github} & C, I, N, T & Fashion & Over 1M outfits, 583K items, and 280M clicks from Alibaba’s iFashion; outfit generation \\
FashionVC  & \href{https://drive.google.com/file/d/1d72E3p4w280-vdCKfXtXZLMULpIScRRR/view}{drive} & I, T & Fashion & \cite{neurostylist} \\
Polyvore & \href{https://github.com/xthan/polyvore-dataset}{github} & I, T & Fashion & \cite{learningfashioncompatibility} \\
IQON3000 & \href{https://anonymity2019.wixsite.com/gp-bpr}{Link} & I, T & Fashion & \cite{GP_BPR} \\
Tianmao & \href{https://tianchi.aliyun.com/dataset/43}{Link} & I, T & Fashion & ~ \\
Taobao & \href{https://tianchi.aliyun.com/dataset/52}{Link} & I, T & Fashion & ~ \\
Free Airports Europe & \href{https://ec.europa.eu/eurostat/databrowser/view/avia_tf_apal/default/table?lang=en}{Link:scraping} & G, Tab & Flights & free airport datasets (\cite{struc2vec}) \\
Free Airports Brazil & \href{https://paperswithcode.com/dataset/brazil-air-traffic}{paperswithcode} & G & Flights & free airport datasets (\cite{struc2vec}) \\
Free Airports USA & \href{https://www.bts.gov/newsroom/may-2016-us-airline-traffic-data}{Link: scraping} & G, Tab & Flights & free airport datasets (\cite{struc2vec}) \\
Allrecipes & \href{https://www.kaggle.com/datasets/elisaxxygao/foodrecsysv1}{kaggle} & I, T & Food & \cite{allrecipes} \\
Recipe & \href{https://www.kaggle.com/datasets/shuyangli94/food-com-recipes-and-user-interactions}{kaggle} & C, T, Tab & Food & From Food.com (\cite{recipe1m}) \\
Dianping  & \href{https://github.com/Meituan-Dianping/ASAP}{Link} & C, T & Food & Aspect sentiment annotations over 18 categories; 5-star ratings; 46,730 reviews (\cite{dianping}) \\
FoodRec & \href{https://acmmultimedia.wixsite.com/foodrec}{Link} & I, T & Food & \cite{Market2Dish} \\
Steam & \href{https://www.kaggle.com/datasets/antonkozyriev/game-recommendations-on-steam}{kaggle} & C, T & Games & \cite{SASRec} \\
PixelRec & \href{https://github.com/westlake-repl/PixelRec}{github} & B, I, T & Image Recommendation & 400K images, extracted text/image features, and multi-scale user–item. New: \cite{PixelRec} \\
MovieLens & \href{https://grouplens.org/datasets/movielens/}{grouplens} & I, T & Movies & Various versions used (e.g., 1M, 100k, 20M) \href{https://dl.acm.org/doi/10.1145/2827872}{link} \\
MovieLens-1M & \href{https://grouplens.org/datasets/movielens/1m}{Link} & C, N & Movies & 1M ratings from 6K users on 4K movies; includes metadata; commonly used for sequence modelling. \\
\bottomrule
\end{tabularx}
}
\caption{Overview of multimodal datasets used in LLM-based recommender systems. \emph{Amazon Various} refers to Music, Clothing, Beauty, Toys, Sports, Games, Baby, Video Games, Electronics, Movies, Books, Fashion and Product Review.\\
\textbf{Modalities}: T = Text, C = Categorical, I = Image, A = Acoustic/Audio, V = Video, G = Graph, Tab = Tabular, POI = Point of Interest, N = Numerical, B = Behavioral}
\label{tab:datasets1}
\end{table*}

\begin{table*}[ht]
\centering
\small
\resizebox{\textwidth}{!}{%
  
}
\caption{Continuation of the Datasets table.\\
\textbf{Modalities}: T = Text, C = Categorical, I = Image, A = Acoustic/Audio, V = Video, G = Graph, Tab = Tabular, POI = Point of Interest, N = Numerical, B = Behavioral}
\label{tab:datasets2}
\end{table*}

\newpage
\subsection{Evaluation Metrics}
\label{sec:metrics}
Evaluating LLM-based Multimodal Recommender Systems (MRS) requires integrating traditional recommendation metrics with those suited for language generation, multimodal fusion, and user-centric evaluation. In this section, we categorise and review the metrics used across recent works, since some previous works concentrate only on the RS metrics (\cite{MRS2_survey}), or they excluded (\cite{MRS1_survey}).

\subsubsection{Recommendation Metrics}
These metrics originate from traditional recommender systems and remain central to evaluating the relevance and ranking quality of recommended items. The most widely used metrics include:

\begin{itemize}
    \item \textbf{Precision@K}, \textbf{Recall@K}, \textbf{F1} and \textbf{Accuracy}: These measure the correctness of the top-K predictions. For example, Recall@K reflects how many relevant items appear in the top-K list. These metrics appear in many papers: They are commonly used in works such as P@K (\cite{LLM4KGC2023, LGIR, LLM_Rec, LLAMA_E, ADKGD, IndustryScopeGPT2024}), R@k (\cite{CrossCBR2022, RLMRec2024, MISSRec, MMContrastive2021, RecInDial, KRagRec, Molar, Mender, ViML}), F$1$ (\cite{PIE, LLAMA_E, IndustryScopeGPT2024}), and Accuracy  (\cite{LLM4KGC2023, RecInDial, HiRMed, KRagRec})
    \item \textbf{Hit Rate (HR@K)}: Measures whether at least one relevant item appears in the top-K list(\cite{EUIRec, PROMO, TripletFusion2024, HLLM2024, PAD2024, DICER2020, HCGCN2022}).
    \item \textbf{NDCG@K} (Normalised Discounted Cumulative Gain): Accounts for the position of relevant items in the ranked list (\cite{HKFR, Molar, NegGen, PatchRec, EUIRec, CrossCBR2022}).
        \begin{align}
        \text{DCG@K} &= \sum_{i=1}^{K} \frac{2^{rel_i} - 1}{\log_2(i+1)} \\
        \text{NDCG@K} &= \frac{\text{DCG@K}}{\text{IDCG@K}}
        \end{align}
        Where \( rel_i \) is the relevance score at rank \( i \), and IDCG is the ideal DCG.
    \item \textbf{MRR} (Mean Reciprocal Rank): Evaluates the rank of the first relevant item for each user (\cite{LGIR, UniMP, MM_Rec, LLLM_Rec, EUIRec, MMGCL2022, TTDS2024}).
        \begin{equation}
        \text{MRR@K} = \frac{1}{|U|} \sum_{u \in U} \frac{1}{\text{rank}_u}
        \end{equation}
        Where \( \text{rank}_u \) is the position of the first relevant item for user \( u \).
    \item \textbf{MAP} (Mean Average Precision): Computes the average precision per user over the top-K results, less frequently used but useful in ranking-sensitive setups (\cite{LGIR, KAR}).
        \begin{align}
        AP_u@K &= \frac{1}{|R_u|} \sum_{k=1}^{K} \text{Precision@k} \cdot \mathbbm{1}[\text{item}_k \in R_u] \\
        MAP@K &= \frac{1}{|U|} \sum_{u \in U} AP_u@K
        \end{align}
    \item \textbf{LogLoss (Binary Cross-Entropy)}: Captures the deviation between predicted probabilities and binary ground-truth labels (\cite{KAR, GIRL, SampleLLM, AltFS, CTRL2023}). Often paired with AUC to evaluate both confidence and ranking:
        \begin{equation}
        \text{LogLoss} = -\frac{1}{N} \sum_{i=1}^{N} \left[ y_i \log \hat{y}_i + (1 - y_i) \log (1 - \hat{y}_i) \right]
        \end{equation}
\end{itemize}

While most of the above are top-K metrics, some others focus more directly on ranking and prediction confidence:
\begin{itemize}
    \item \textbf{AUC} (Area Under the Curve): Measures the probability that a positive item is ranked above a negative one (\cite{HLLM2024, CDR2021, CTRL2023, SampleLLM, AltFS, MM_Rec}).
        \begin{equation}
        \text{AUC} = \frac{1}{|P||N|} \sum_{p \in P} \sum_{n \in N} \mathbb{1}[s(p) > s(n)]
        \end{equation}
        where \( P \) and \( N \) are sets of positive and negative items, and \( s(\cdot) \) is the predicted score.
    \item \textbf{Group AUC (GAUC)} (\cite{TaobaoMRS_2024}) Computes AUC separately for each user (or group) and then averages the result, weighted by impression count:
        \begin{equation}
        \text{GAUC} = \frac{\sum_{u} w_u \cdot \text{AUC}_u}{\sum_{u} w_u}
        \end{equation}
    \item \textbf{Relative Improvement (RelaImpr)}: Quantifies AUC improvement over a baseline, normalised by the AUC gap from random chance (0.5) (\cite{CTRL2023, CDR2021}).
        \begin{equation}
        \text{RelaImpr} = \left( \frac{\text{AUC}_{\text{model}} - 0.5}{\text{AUC}_{\text{base}} - 0.5} - 1 \right) \times 100\%
        \end{equation}
\end{itemize}

These metrics are widely adopted in multimodal RS works such as MMGCN~\cite{wei2019mmgcn}, FREEDOM~\cite{zhou2023freedom}, and DualGNN~\cite{wang2021dualgnn}. However, they assume clearly defined ground-truth relevance labels and do not assess language quality or multimodal integration. One additional metric is \textbf{Median Rank}, used in \cite{ViML}, which reports the median rank of the first relevant item across users.

\subsubsection{Language Generation n-gram Metrics}
Text quality becomes essential when LLMs generate reviews, explanations, item summaries, or full user journeys. A set of standard n-gram-based metrics is widely used for this purpose:

\textbf{BLEU}~\cite{bleu2002} measures the geometric mean of n-gram precisions, with a brevity penalty to discourage overly short outputs:
\begin{equation}
\text{BLEU} = \text{BP} \cdot \exp\left( \sum_{n=1}^N w_n \log p_n \right)
\end{equation}
where \( p_n \) is the modified precision for n-grams, \( w_n \) are typically uniform weights, and BP is the brevity penalty. Widely used in~\cite{IntellectSeeker, LLAMA_E, UniMP, PersonalizedPromptLearning, RecInDial, NOR2019}.

\textbf{ROUGE-N}~\cite{rouge2004} measures the recall of overlapping n-grams between generated and reference text:
\begin{equation}
\text{ROUGE-N} = \frac{\# \text{matched n-grams}}{\# \text{reference n-grams}}
\end{equation}
Frequently reported in~\cite{VIP5, P52023, PersonalizedPromptLearning}.

\textbf{ROUGE-L}~\cite{rouge2004} uses the length of the longest common subsequence (LCS) between generated and reference texts:
\begin{equation}
\text{ROUGE-L} = \frac{LCS(\text{gen}, \text{ref})}{\text{len}(\text{ref})}
\end{equation}
It captures sequence-level overlap and is normalised by reference length.

\textbf{METEOR}~\cite{meteor2005} computes an F1 score over unigram matches, with a penalty for fragmentation:
\begin{equation}
\text{METEOR} = F_{\text{mean}} \cdot (1 - \text{Penalty})
\end{equation}
where \( F_{\text{mean}} \) is the harmonic mean of precision and recall. Less common, used in~\cite{IntellectSeeker, UniMP, LLAMA_E}.

\textbf{Structured Generation Metrics}: Variants like \textbf{G-BLEU} and \textbf{G-ROUGE} apply n-gram logic to structured outputs (e.g., KG triples treated as sentences)~\cite{GraphJudger}.

\textbf{Valid Ratio}: Reports the proportion of syntactically or structurally valid outputs (e.g., well-formed triples, valid SQL)~\cite{TripletFusion2024}.

\subsubsection{Language Generation Semantic Metrics}
Embedding-based metrics have emerged with transformer models to assess semantic similarity beyond surface-level overlap:

\textbf{BERTScore}~\cite{bertscore2020} uses contextual embeddings from a pretrained BERT model:
\begin{align}
\text{Precision} &= \frac{1}{|X|} \sum_{x \in X} \max_{y \in Y} \cos(e_x, e_y) \\
\text{Recall} &= \frac{1}{|Y|} \sum_{y \in Y} \max_{x \in X} \cos(e_y, e_x) \\
\text{BERTScore} &= \frac{2 \cdot \text{Precision} \cdot \text{Recall}}{\text{Precision} + \text{Recall}}
\end{align}
Used in~\cite{GraphJudger, UniMP, LLAMA_E}.

\textbf{BLEURT}~\cite{bleurt2020} is a learned metric combining pretrained contextual embeddings with fine-tuning to predict human judgments. Found in~\cite{ICPC}.

\textbf{Perplexity (PPL)} used in ~\cite{RecInDial, LLAMA_E}, evaluates fluency via inverse likelihood under a language model:
\begin{equation}
\text{PPL} = \exp\left(-\frac{1}{N} \sum_{i=1}^N \log P(w_i | w_{<i})\right)
\end{equation}

Though these metrics support explainable RS, review generation, and dialogue settings, they may overlook diversity and factual correctness.

\textbf{Factual Alignment Metrics}: Task-specific scores such as Feature Matching Ratio (FMR), Feature Coverage Ratio (FCR), and Feature Diversity (DIV)~\cite{features_metrics} assess whether generated explanations accurately reflect item attributes. These are still rarely used~\cite{PersonalizedPromptLearning}.

\subsubsection{Agent and Simulation Metrics}
In interactive or simulated environments (e.g., dialogue systems, journey planners), LLMs-as-agents require different metrics:

\textbf{RL-based Metrics} (e.g., iALP~\cite{iALP}):
\begin{itemize}
    \item \textbf{Return}: Total reward accumulated per recommendation episode~\cite{return_metric}.
    \item \textbf{Trajectory Length}: Duration of user engagement within a session~\cite{trajectory_length_metric}.
    \item \textbf{Average Reward}: Mean reward per action, reflecting policy effectiveness~\cite{average_reward_metric}.
\end{itemize}
These vary across simulations and are sometimes complemented by A/B tests or online evaluations.

\paragraph{Preference-Following Axes}
Mender~\cite{Mender} proposes five axes to evaluate preference-following behavior of a generation:
(1) \textit{Preference-based Recommendation},
(2) \textit{Fine- and Coarse-grained Steering},
(3) \textit{Sentiment Following},
(4) \textit{History Consolidation}, and
(5) \textit{Steering Accuracy} measured via combined hit-rate metrics like \( m@k \).
These aim to assess controllability and user alignment in generative recommenders.



\subsubsection{Multimodal Evaluation}
Evaluating the effectiveness of multimodal fusion requires assessing whether textual, visual, and other modalities are meaningfully combined. We include here metrics often used to assess the distinct contribution or alignment of multiple modalities, even if some originated in unimodal RS. Relevant metrics include:
\begin{itemize}
    \item \textbf{CLIPScore}~\cite{clipscore}: A reference-free metric that uses CLIP's image-text embeddings to assess the alignment between generated textual explanations and visual content, without relying on human-written references. Used in multimodal explanation generation tasks~\cite{personalized_showcases}.
    \item \textbf{Catalog Coverage} (\cite{catalog_coverage}): represents the fraction of catalog items that appeared in at least one top-n recommendation list of the users in the test set. These metrics are commonly used in general or sequential RS to evaluate diversity and coverage, as in~\cite{bias, LLLM_Rec}.
    \item \textbf{Serendipity}(\cite{serendipity}): measures the average number of correct recommendations for each user that are not recommended by a popularity baseline.
    \item \textbf{Novelty}(\cite{novelty_metric}): computes the negative log of the relative item popularity, or self-information.
\end{itemize}

\subsubsection{Human and LLM Evaluation}
Manual or LLM-based evaluations are increasingly used to assess the subjective quality of LLM-based outputs:

\begin{itemize}
    \item \textbf{Fluency, Relevance, Diversity}: Rated by human annotators, often via crowdsourcing. Some works also report automatic diversity metrics such as distinct-n~\cite{RecInDial}.
    \item \textbf{Polarisation}: Captures user reactions or self-reported shifts in sentiment or preference~\cite{RecSysLLMsP}.
    \item \textbf{Popularity}: Measures the distribution and frequency of tags across similar objects to detect spammy or overly generic labels~\cite{TagGPT}.
    \item \textbf{Practicality}: Assesses informativeness and efficiency of tag sets via least effort (tag count per object) and facet coverage (semantic breadth)~\cite{TagGPT}.
    \item \textbf{Uniformity}: Quantifies semantic redundancy among tags to detect divergence or noise in tagging behavior~\cite{TagGPT}.
    
    \item \textbf{Qualitative Journey Coherence}: Involves manual inspection of extracted user journeys to assess coherence and interpretability~\cite{ICPC}.
    \item \textbf{User Studies and A/B Tests}: Used to compare variants of generated journeys, names, or recommendations~\cite{PROMO, KAR}.
    \item \textbf{Domain-specific Metrics}: For instance, \textbf{Clinical Relevance Score (CRS)} rates medical appropriateness on a 1–5 scale, while \textbf{Coverage Rate (CR)} measures the proportion of relevant diagnostic tests included~\cite{HiRMed}.
    \item \textbf{LLM-based Evaluation (Win/Tie/Lose, Advantage)}: Some works substitute human judges with LLMs like ChatGPT~\cite{GIRL}, comparing outputs from different systems in terms of detail, relevance, and conciseness. Metrics such as \textit{Win Rate}, \textit{Tie Rate}, \textit{Lose Rate}, and \textit{Advantage} (Win minus Lose) are computed from these judgments.
\end{itemize}

These evaluations provide deep insight into real-world utility. While human-based assessments are costly and difficult to scale, LLM-based evaluations offer a practical and increasingly adopted alternative for subjective quality assessment.

\begin{itemize}
    \item \textbf{Click-through Rate (CTR)}: Measures the proportion of users who click on a recommended item. It serves as a primary indicator of immediate recommendation effectiveness~\cite{HKFR, TaobaoMRS_2024, SampleLLM, AltFS, CDR2021}.
    \item \textbf{Gross Merchandise Value (GMV)}: Represents the total monetary value of transactions resulting from the recommender system, reflecting its commercial impact~\cite{TripletFusion2024}.
    \item \textbf{User Interaction Count (likes, shares, comments)}: Captures deeper forms of engagement beyond clicks, such as video play time, likes, and collections~\cite{PROMO}.
\end{itemize}

\newpage
\subsection{Notation}
\label{sec:notations}
The corpus of this paper contains many terms from NLP and RS. We list all of them in one table for simplicity and direct reference.
\begin{table}[htbp]
    \centering
    \small
    \caption{Notation}
    \label{tab:notation1}
    \begin{tabularx}{\columnwidth}{@{}>{\raggedright\arraybackslash}X@{}}
    \toprule
    \textbf{Pretrained Language Models / LLM Families} \\ \addlinespace
    \toprule
    \textbf{Natural Language Processing (NLP)}: The broader domain includes LLM reasoning and structured prompt design techniques. \\ \addlinespace
    \textbf{Large Language Model (LLM)}: Advanced AI models designed to understand and generate human-like text based on extensive training data, used for tasks like text generation and question-answering. \\ \addlinespace
    \textbf{Large Language Model (VLM)}: Visual Language model. \\ \addlinespace
    \textbf{Pre-trained Language Model (PLM)}: A language model trained on large text corpora, ready to be fine-tuned for specific tasks. \\ \addlinespace 
    \textbf{Generative Pre-trained Transformer (GPT)}: A family of LLMs including GPT-3, used in various prompting-based approaches. \\ \addlinespace
    \textbf{Generative Pre-trained Transformer 2 (GPT-2)}: Second-generation GPT model by OpenAI; used in hybrid prompting examples. \\ \addlinespace
    \textbf{Large Language Model Meta AI (LLAMA)}: Open LLM by Meta. \\ \addlinespace
    \textbf{Pathways Language Model (PaLM)}: LLM developed by Google. \\ \addlinespace
    \textbf{Language Model for Dialogue Applications (LaMDA)}: Google's conversational LLM, used in ICPC. \\ \addlinespace
    \textbf{Bidirectional Encoder Representations from Transformers (BERT)}: A commonly used PLM; mentioned in hybrid prompting. \\ \addlinespace
    \textbf{Robustly Optimized BERT Pretraining Approach (RoBERTa)}: A transformer-based PLM used for text encoding. \\ \addlinespace
    \textbf{Text-To-Text Transfer Transformer (T5)}: A unified transformer model that frames all NLP tasks (e.g., translation, summarisation) as text-to-text problems. \\ \addlinespace
    \textbf{Sentence-BERT (SBERT)}: Sentence-level embedding model based on BERT. \\ \addlinespace
    \textbf{Vision-and-Language BERT (ViLBERT)}: Multimodal transformer architecture for jointly modeling image and text. \\ \addlinespace
    \toprule
    \textbf{Model Architectures and Components} \\ \addlinespace
    \toprule
    \textbf{Graph Convolutional Network (GCN)}: Used to encode subgraphs or structural dependencies in KG-based methods or collaborative graphs. \\ \addlinespace
    \textbf{Graph Neural Network (GNN)}: Used for structural learning on graphs, e.g., in KG-based recommendation. \\ \addlinespace
    \textbf{Multilayer Perceptron (MLP)}: A neural network with multiple fully connected layers that models complex patterns. \\ \addlinespace 
    \textbf{Convolutional Neural Network (CNN)}: A deep network using convolutional layers to process and analyse grid-like data, such as images. \\ \addlinespace
    \textbf{Generative Adversarial Network (GAN)}: A generator and discriminator framework used for post-hoc embedding refinement. \\ \addlinespace
    \textbf{Long Short-Term Memory (LSTM)}: A type of recurrent neural network used to generate dynamic soft prompts in EUIRec. \\ \addlinespace
    \textbf{Bidirectional Long Short-Term Memory (Bi-LSTM)}: Recurrent model encoding sequences in both directions; used for dual-view encoding. \\ \addlinespace
    \textbf{Mixture-of-Experts (MoE)}: A neural network architecture combining several expert models through a gating mechanism. \\ \addlinespace
    \textbf{Vision Transformer (ViT)}: Used for encoding image content in zero-tuning setups. \\ \addlinespace
    \textbf{Variational Autoencoder (VAE)}: A generative model often used for latent representation learning. \\ \addlinespace
    \textbf{Residual Quantized Variational Autoencoder (RQ-VAE)}: A variant of VAE used for quantising visual inputs into semantic IDs. \\ \addlinespace
    \textbf{Beta-Variational Autoencoder ($\beta$-VAE)}: A VAE variant where the KL divergence term is scaled by a factor $\beta$ to encourage more disentangled latent representations. \\ \addlinespace
    \textbf{Relational Graph Convolutional Network (R-GCN)}: A type of GCN that incorporates relation types (edges) when propagating node information; used for encoding entities in knowledge graphs. \\ \addlinespace
    \bottomrule
    \end{tabularx}
\end{table}

\begin{table}[htbp]
    \centering
    \small
    \caption{Notation}
    \label{tab:notation2}
    \begin{tabularx}{\columnwidth}{@{}>{\raggedright\arraybackslash}X@{}}
    \toprule
    \textbf{Training Objectives and Techniques} \\ \addlinespace
    \toprule
    \textbf{Contrastive Learning (CL)}: A self-supervised method that distinguishes similar from dissimilar data pairs. \\ \addlinespace 
    \textbf{Masked Language Modelling (MLM)}: A training objective involving token masking, adapted for multimodal alignment and pretraining. \\ \addlinespace
    \textbf{Next Sentence Prediction (NSP)}: A training task for PLMs like BERT, predicting whether two sentences follow each other in a document. \\ \addlinespace
    \textbf{Info Noise Contrastive Estimation (InfoNCE)}: A contrastive loss used to train representations by maximising agreement between positive pairs. \\ \addlinespace
    \textbf{Kullback–Leibler divergence (KL)}: A measure of divergence between two probability distributions; used in VAE regularisation to enforce latent independence. \\ \addlinespace
    \textbf{Maximum Mean Discrepancy (MMD)}: A kernel-based statistical distance used to compare distributions; applied in PAD to align collaborative and semantic experts. \\ \addlinespace
    \textbf{Semantic-Aware Contrastive Learning (SCL)}: A contrastive learning variant that encourages semantic consistency across modalities while maintaining their separation. \\ \addlinespace
    \textbf{Fine-Tuning (FT)}: Adjusting a pre-trained model for specific tasks using techniques like \textbf{Adapters} (small modules inserted into the model) and \textbf{LoRA} (Low-Rank Adaptation) to efficiently update parameters without modifying the entire model. \\ \addlinespace
    \textbf{Low-Rank Adaptation (LoRA)}: Lightweight fine-tuning method that adapts a small subset of model parameters via low-rank decomposition. \\ \addlinespace
    \textbf{Quantised Low-Rank Adaptation (QLoRA)}: Extension of LoRA using quantised weights to reduce memory and computational requirements. \\ \addlinespace
    \textbf{Instruction Tuning (IT)}: Fine-tuning strategy where LLMs are trained on prompt-response pairs. \\ \addlinespace
    \textbf{Supervised Fine-Tuning (SFT)}: Fine-tuning approach using labelled instruction–response data. \\ \addlinespace
    \textbf{Reinforcement Learning from Human Feedback (RLHF)}: Training paradigm used to align LLMs with human preferences, significantly improving conversational abilities. \\ \addlinespace
    \textbf{Reinforcement Learning with a Reward Function (RLRF)}: Variant of RLHF focused on reflective/self-generated feedback. A training strategy that fine-tunes a language model using reinforcement learning (e.g., PPO), guided by a reward model that scores the quality of generated outputs. \\ \addlinespace 
    \textbf{Proximal Policy Optimisation (PPO)}: RL algorithm for fine-tuning with human/recruiter feedback. \\ \addlinespace
    \toprule
    \textbf{Prompting and Reasoning Techniques} \\ \addlinespace
    \toprule
    \textbf{Chain-of-Thought (CoT)}: A reasoning technique where the model generates intermediate steps to arrive at a final answer, enhancing problem-solving and interpretability. \\ \addlinespace
    \textbf{In-Context Learning (ICL)}: A method where the model is provided with examples or additional context within the input prompt to guide its responses without explicit retraining. \\ \addlinespace
    \textbf{Reasoning-Prompting Techniques}: Strategies that guide the model’s reasoning process through carefully designed prompts or instructions, improving logical deductions and response accuracy. \\ \addlinespace
    \textbf{Double Checking (DC)}: A reasoning enhancement method where an LLM validates or verifies its predictions. \\ \addlinespace
    \textbf{Direct Prompt Methods}: Approaches where the model is given straightforward prompts without additional reasoning steps or external data integration, offering simplicity and speed but less depth. \\ \addlinespace
    \toprule
    \textbf{System-Level / Agents / Planning} \\ \addlinespace
    \toprule
    \textbf{LLM Agent}: Autonomous system powered by a Large Language Model that interprets and responds to user queries with coherent and contextually relevant information. The LLM Agent leverages a comprehensive knowledge graph to access structured data and employs advanced reasoning-prompting techniques to generate accurate and reliable query responses. \\ \addlinespace
    \textbf{REACT}: A framework that combines reasoning and acting, allowing the model to perform actions like querying databases as part of its response generation process. \\ \addlinespace
    \textbf{Monte Carlo Tree Search (MCTS)}: A decision-making algorithm that explores action spaces by building a search tree guided by random simulations and reward estimates, balancing exploration and exploitation to find optimal outcomes. \\ \addlinespace
    \textbf{Upper Confidence Bounds for Trees (UCT)}: A decision-making algorithm used in Monte Carlo Tree Search balances exploration and exploitation by selecting tree nodes based on their average reward and visit count. UCT helps explore large decision spaces efficiently by prioritising nodes with high potential but limited visits.\\ \addlinespace
    \bottomrule
    \end{tabularx}
\end{table}

\begin{table}[htbp]
    \centering
    \small
    \caption{Notation}
    \label{tab:notation3}
    \begin{tabularx}{\columnwidth}{@{}>{\raggedright\arraybackslash}X@{}}
    \toprule
    \textbf{Retrieval $\&$ Semantic Tools} \\ \addlinespace
    \toprule
    \textbf{Retrieval-Augmented Generation (RAG)}: LLM enhancement method using external document retrieval. \\ \addlinespace
    \textbf{Facebook AI Similarity Search (FAISS)}: A library for efficient similarity search, used for semantic retrieval in agent pipelines. \\ \addlinespace
    \textbf{Large Language Model to Vector (LLM2Vec)}: Technique that turns LLMs into encoders for downstream alignment tasks. \\ \addlinespace
    \toprule
    \textbf{Multimodal and Fusion Concepts} \\ \addlinespace
    \toprule
    \textbf{Multimodal Large Language Model (MLLM)}: Refers to LLMs adapted to handle multimodal inputs like image and text. \\ \addlinespace
    \textbf{Contrastive Language–Image Pretraining (CLIP)}: A vision-language model used for encoding image features. \\ \addlinespace
    \textbf{Optical Character Recognition (OCR)}: Extracts text from images for tagging pipelines with LLMs. \\ \addlinespace
    \textbf{Automatic Speech Recognition (ASR)}: Extracts text from audio, often used in multimodal tag extraction with LLMs. \\ \addlinespace
    \textbf{JavaScript Object Notation (JSON)}: Human-readable structured data format, used for code-like prompts. \\ \addlinespace
    \textbf{Region of Interest (ROI)}: Typically refers to image regions, used in visual recommendation inputs. \\ \addlinespace
    \toprule
    \textbf{Recommender Systems} \\ \addlinespace
    \toprule
    \textbf{Multimodal Recommender System (MRS)}: Recommender systems that integrate multiple modalities (text, image, etc.). \\ \addlinespace
    \textbf{Recommender System (RS)}: General term for systems that make item suggestions to users. \\ \addlinespace
    \textbf{Knowledge Graph (KG)}: A structured representation of information where entities are interconnected through relationships, facilitating complex queries and reasoning. \\ \addlinespace
    \textbf{Identifier (ID)}: Unique tokens (e.g., user or item IDs) used in collaborative filtering or to anchor structured representations. \\ \addlinespace
    \textbf{Multi-modal Interest Discovery (MID)}: A module that clusters item features to generate interpretable user interest prototypes across modalities. \\ \addlinespace
    \textbf{Semantic Extraction Network (SEN)}: A module designed to extract fine-grained, semantically rich local features from multimodal inputs. \\ \addlinespace
    \textbf{Graph-based User–Item graph (\GUI)}: Refers to interaction graphs capturing user–item relations, used in soft prompting for input representation. \\ \addlinespace
    \toprule
    \textbf{Domains / Inputs / Context} \\ \addlinespace
    \toprule
    \textbf{Global Navigation Satellite System (GNSS)}: In Multi-LLM-GNSS, refers to the domain where LLMs are applied. \\ \addlinespace
    \textbf{Point of Interest (POI)}: Geospatial points used in location-based recommendation, typically mapped as latitude-longitude entries. \\ \addlinespace
    \textbf{Industrial Park Planning and Operation (IPPO)}: The design and management of designated industrial areas, focusing on infrastructure, zoning, and resource allocation to support businesses and economic growth. \\ \addlinespace
    \toprule
    \textbf{Attention} \\ \addlinespace
    \toprule
    \textbf{Self-Attention}: Allows each element in a sequence to focus on all other elements, capturing internal dependencies. \\ \addlinespace
    \textbf{Multi-Head Attention (MHA)}: Utilises multiple attention heads to simultaneously capture diverse aspects of relationships within the data. \\ \addlinespace
    \textbf{Bahdanau/Additive Attention}: Calculates attention scores using a feedforward network to combine queries and keys additively. \\ \addlinespace
    \textbf{Luong/Multiplicative Attention}: A variant that computes attention scores using methods like dot, general, or concatenation, primarily used in sequence-to-sequence models. \\ \addlinespace
    \textbf{Cross-Attention}: Enables one sequence to attend to another, facilitating interactions between different input sources. \\ \addlinespace
    \textbf{Co-Attention} extends the attention mechanism to handle interactions between two distinct data sources or sequences. Unlike cross-attention, which is typically unidirectional (one sequence attends to another), co-attention is bidirectional, enabling both sequences to simultaneously influence each other’s representations. \\ \addlinespace
    \bottomrule
    \end{tabularx}
\end{table}

\clearpage
\end{document}